\documentclass[twocolumn,preprintnumbers,amsmath,amssymb,superscriptaddress,pre,showpacs]{revtex4}
\usepackage[dvips]{graphicx}
\usepackage{dcolumn}
\usepackage{bm}
\usepackage[dvips]{epsfig}

\begin{document}

\title{Stress overshoot in a simple yield stress fluid:\\ an extensive study combining rheology and velocimetry}

\author{Thibaut Divoux}
\affiliation{Universit\'e de Lyon, Laboratoire de Physique, \'Ecole Normale Sup\'erieure de
Lyon, CNRS UMR 5672, 46 All\'ee d'Italie, 69364 Lyon cedex 07, France.}
\author{Catherine Barentin}
\affiliation{Laboratoire de Physique de la Mati\`ere Condens\'ee et Nanostructures, Universit\'e de Lyon; Universit\'e Claude Bernard Lyon I, CNRS UMR 5586 - 43 Boulevard du 11 Novembre 1918, 69622 Villeurbanne cedex, France.}
\author{S\'ebastien Manneville}
\affiliation{Universit\'e de Lyon, Laboratoire de Physique, \'Ecole Normale Sup\'erieure de
Lyon, CNRS UMR 5672, 46 All\'ee d'Italie, 69364 Lyon cedex 07, France.}
\affiliation{Institut Universitaire de France.}
\date{\today}

\begin{abstract}
We report a large amount of experimental data on the stress overshoot phenomenon which takes place during start-up shear flows in a simple yield stress fluid, namely a carbopol microgel. A combination of classical rheological measurements and ultrasonic velocimetry makes it possible to get physical insights on the transient dynamics of both the stress $\sigma(t)$ and the velocity field across the gap of a rough cylindrical Couette cell during the start-up of shear under an applied shear rate $\dot\gamma$. (i) At small strains ($\gamma <1$), $\sigma(t)$ increases linearly and the microgel undergoes homogeneous deformation. (ii) At a time $t_m$, the stress reaches a maximum value $\sigma_m$ which corresponds to the failure of the microgel and to the nucleation of a thin lubrication layer at the moving wall. (iii) The microgel then experiences a strong elastic recoil and enters a regime of total wall slip while the stress slowly decreases. (iv) Total wall slip gives way to a transient shear-banding phenomenon, which occurs on timescales much longer than that of the stress overshoot and has been described elsewhere [Divoux \textit{et al., Phys. Rev. Lett.}, 2010, \textbf{104}, 208301]. This whole sequence is very robust to concentration changes in the explored range ($0.5 \le C \le 3 \%$ w/w). We further demonstrate that the maximum stress $\sigma_m$ and the corresponding strain $\gamma_m=\dot\gamma t_m$ both depend on the applied shear rate $\dot \gamma$ and on the waiting time $t_w$ between preshear and shear start-up: they remain roughly constant as long as $\dot\gamma$ is smaller than some critical shear rate $\dot\gamma_w\sim 1/t_w$ and they increase as weak power laws of $\dot \gamma$ for  $\dot\gamma> \dot\gamma_w$. Finally, by changing the boundary conditions from rough to smooth, we show that there exists a critical shear rate $\dot \gamma_s$ fixed by the wall surface roughness below which slip at both walls allows for faster stress relaxation and for stress fluctuations strongly reminiscent of stick-slip. Interestingly, the value of $\dot \gamma_s$ is observed to coincide with the shear rate below which the flow curve displays a kink attributed to wall slip.
\end{abstract}
\maketitle

The transient response of complex materials to the application of an external shear deformation is of huge importance not only for the practical use of such materials but also during the processing stage. The archetypal experiment used for transient rheological characterization is a ``start-up experiment'' where a constant shear rate $\dot \gamma$ is applied from rest and the subsequent stress response is monitored until steady-state is reached. A host of widely different systems from soft and hard condensed matter have been reported to present a non-monotonic stress response during start-up experiments. Rough\-ly, the stress $\sigma$ versus time $t$ first increases linearly, reaches a maximum value denoted $\sigma_m$ at a time $t_m$ and then decreases towards it steady-state value. This temporal sequence is usually referred to as a {\it stress overshoot}. It has been reported experimentally for both amorphous materials, such as amorphous polymers \cite{Arruda:1995,Melick:2003,Melick:2003b} and metallic glasses \cite{Aken:2000,Johnson:2002}, and for soft glassy materials, na\-me\-ly emulsions \cite{Partal:1999,Becu:2005,Becu:2006,Ovarlez:2008}, foams \cite{Khan:1988,Raufaste:2010}, microgels \cite{Islam:2004,Coussot:2009}, both dry \cite{Wroth:1958} and im\-mer\-sed \cite{Geminard:1999} granular materials, organoclays nanocomposites \cite{Letwimolnun:2007}, and colloidal suspensions \cite{Nguyen:1983,Persello:1994,Derec:2003,Mahaut:2008,Koumakis:2011}. Numerical models of such systems, e.g. bidisperse Lennard-Jones glasses \cite{Rottler:2003,Varnik:2004,Rottler:2005}, mesoscopic models of amorphous materials \cite{Langer:2003,Jagla:2007,Jagla:2010} (for a review, see also Ref.~\cite{Falk:2010pp}), models of soft glasses \cite{Sollich:1998,Fielding:2000,Derec:2003,Moorcroft:2011}, brownian dynamics simulations of particulate gels \cite{Whittle:1997} as well as molecular dynamics simulations \cite{Albano:2004,Xu:2006} and mode coupling theory \cite{Zausch:2008}, also predict a stress overshoot.

The above materials are all {\it yield stress} materials, i.e. they share the property of turning from solidlike at rest and at low shear stresses to liquidlike above a characteristic shear stress known as the {\it yield stress}. Therefore, one may wonder whether the stress overshoot phenomenon is the hallmark of the existence of a yield stress and how it may be related to the shear-induced solid-to-liquid transition. In fact, stress overshoots are also commonly found in viscoelastic fluids that do not present a yield stress, such as wormlike micelle solutions \cite{Berret:1997b,Lerouge:1998,Soltero:1999,Lerouge:2000,Decruppe:2001,Lerouge:2004,Ganapathy:2008}, polymer melts \cite{Yamamoto:2004,Boukany:2009a}, and entangled polymer solutions \cite{Tapadia:2004,Ravindranath:2008b,Boukany:2009b,Boukany:2010}. In this latter case, a possible microscopic interpretation of the stress overshoot has been proposed by Wang {\it et al.}, in which the polymers disentangle after the stress maximum, suggesting a ``yieldlike'' behaviour \cite{Tapadia:2004,Tapadia:2006a,Wang:2009,Boukany:2010}. Experiments based on combined rheology and velocimetry have unveiled the development of heterogeneous flows characterized by wall slip and shear bands after the stress maximum is reached \cite{Ravindranath:2008a,Boukany:2009a,Boukany:2009b}. Molecular imaging has provided even more insights into the interplay between wall slip and polymer stretching, disentangling, and recoil \cite{Boukany:2010}. Moreover, recent numerical calculations of polymeric solutions have also shown that transient shear banding was involved during the stress overshoot \cite{Adams:2009,Adams:2010}.

The current state-of-the art is, however, much less clear in the case of materials with a yield stress. This is probably due both to the wide variety of the microstructures of these materials and to the lack of time-resolved local data on such systems during start-up experiments. Rheology alone only allows for a simple interpretation in which (i) the initial growth of the stress corresponds to elasticlike response, (ii) the stress maximum to yielding, and (iii) the stress decrease towards steady state to fluidlike response \cite{Nguyen:1983,Nguyen:1992}. This oversimplified view obviously lacks local, and ideally microscopic, support. Unfortunately, the issues raised by the existence of a yield stress \cite{Barnes:1999,Moller:2009a} are difficult enough that most previous local studies have dealt with characterizing the steady state, e.g. by asking whether the flow close to yielding is homogeneous or rather displays wall slip and/or shear banding depending on whether the material shows aging and thixotropy or not \cite{Ragouilliaux:2007,Moller:2008}. To the best of our knowledge, only qualitative studies based on direct visualizations are available on the local behaviour of yield stress materials during the stress overshoot \cite{Magnin:1990,Persello:1994,Pignon:1996}. These studies have shown that the initial stage indeed corresponds to homogeneous elastic strain but that strong flow heterogeneity occurs after the stress overshoot in the form of wall slip or bulk fracture. To fully investigate the interplay between shear and microstructure in the short-time response of yield stress fluids, it is clear that more quantitative local measurements are required in the same spirit as for recent progress on entangled polymers.

In this article, we focus on the stress overshoot phe\-no\-me\-non in the case of a {\it simple} yield stress fluid (YSF) while keeping in mind that the present findings may well turn out to be shared by many other glassy systems with similar stress responses. We recall that simple YSF encompass wet foams, emulsions, and microgels \cite{Moller:2009b,Coussot:2010}. These materials are all characterized by a ``jammed'' microstructure constituted of soft, deformable elements (bubbles, droplets, and swollen microgel ``particles'' respectively) compressed together into an amorphous arrangement \cite{Weeks:2007}. Although the rheological behaviour of simple YSF has generated a huge body of literature, very few thorough studies have been devoted to the stress overshoot phenomenon so far. Still understanding the short-time response is crucial to build a general picture of glassy systems under shear. In the case of {\it foams}, the stress overshoot has been reported to occur at a constant characteristic strain ($\gamma_m=\dot\gamma t_m \lesssim1$) and the stress maximum $\sigma_m$ was found to increase with the gas volume fraction, but the dependence of $\sigma_m$ with the imposed shear rate has not been investigated in detail \cite{Khan:1988}. Also, stress overshoots were observed in numerical simulations of foams \cite{Durian:1995,Okuzono:1995,Benito:2008,Barry:2010,Raufaste:2010} and clearly linked to plastic events (``T1'' events) \cite{Kabla:2007} but no systematic study of the overshoot characteristics has been conducted yet. As for {\it emulsions}, both attractive and repulsive systems display stress overshoots \cite{Partal:1999,Becu:2005,Becu:2006,Ovarlez:2008}, but no systematic study has been conducted either. Finally, the overshoot data available on {\it microgels} are also scarce: a stress overshoot has been reported at high shear rates ($\dot \gamma \gtrsim 2$~s$^{-1}$) in carbopol 980 neutralized by TEA \cite{Islam:2004} (see details about the carbopol preparation below) as well as in hair gel solutions composed mainly of carbopol \cite{Coussot:2009} for all imposed shear rates ranging from 2.10$^{-4}$ to 4.10$^{-2}$~s$^{-1}$.

Here we report on an extensive series of start-up experiments performed in carbopol microgels (ETD 2050 neutralized with NaOH). The article is structured as follows. In Section 2, the sample preparation is described in detail. We review the current state-of-the-art on carbopol microgel properties and microstructure. We also describe the experimental techniques used in this study and provide a full rheological characterization of our carbopol samples. In particular, we show that the hollow glass spheres used to seed the microgel and to provide acoustic contrast for ultrasonic measurements have a negligible effect on the viscoelastic properties of carbopol microgels. Section 3 gathers the experimental results obtained for two different geometries. We first report a spatially and temporally resolved study of the stress overshoot in a cylindrical Couette cell, including velocity profiles before and after the stress maximum. We show that (i) at small strains ($\gamma <1$), the microgel undergoes homogeneous deformation; (ii) the maximum stress $\sigma_m$ corresponds to the failure of the microgel and to the nucleation of a thin lubrication layer at the moving wall; (iii) the microgel then experiences a strong elastic recoil and enters a regime of total wall slip while the stress decreases. This whole sequence is very robust to concentration changes in the explored range ($0.5 \le C \le 3 \%$ w/w). We then provide a full characterization of the stress overshoot in the plate-and-plate geometry with respect to the shear rate, to the waiting time $t_w$ between preshear and shear start-up, to the carbopol concentration $C$, and to the boundary conditions (smooth vs rough). We find that the maximum stress $\sigma_m$ and the corresponding strain $\gamma_m=\dot\gamma t_m$ both depend on the applied shear rate $\dot \gamma$ and on the waiting time $t_w$: they remain roughly constant as long as $\dot\gamma$ is smaller than some critical shear rate $\dot\gamma_w\sim 1/t_w$ and they increase as weak power laws of $\dot \gamma$ for  $\dot\gamma> \dot\gamma_w$. Moreover, all the $\sigma_m$ vs $\dot\gamma$ data obtained with rough boundary conditions are shown to collapse well onto a single master curve if one considers $\sigma_m/G_0$ vs $\dot\gamma t_w$, where $G_0$ is the elastic modulus of the microgel. Finally, changing the boundary conditions from rough to smooth leads to important changes in this general scenario: for $\dot \gamma$ lower than some critical shear rate $\dot \gamma_s$ fixed by the value of the wall surface roughness, slip at both walls allows for faster stress relaxation and stress fluctuations strongly reminiscent of stick-slip appear, whereas for $\dot \gamma > \dot \gamma_s$, the stress evolution is very similar to the one described for rough boundary conditions.

\section{Experimental}

\begin{figure}[t]\tt
\centering
\includegraphics[width=0.9\linewidth]{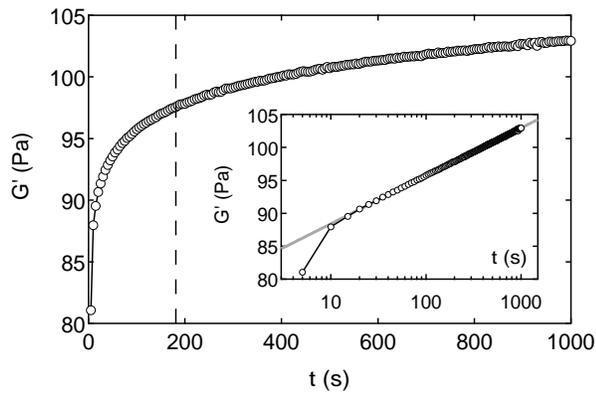}
\caption{Elastic modulus $G'$ as a function of time $t$ after preshear at 1000~s$^{-1}$ for 1 min and at -1000~s$^{-1}$ for 1 min. The oscillation frequency is 1~Hz and the imposed strain amplitude is $\gamma = 1$~\%. The dashed line indicates the waiting time $t_w=180$~s used in the present work (except in Section~\ref{protocol} where $t_w$ is varied). Inset: same data in semilogarithmic scales. The grey line is the best logarithmic fit for $t>10$~s. Experiment performed on a pure carbopol microgel at 1~\% w/w in a plate-and-plate geometry of gap $e=1$~mm under rough boundary conditions (sand paper of roughness 46 $\mu$m).}
\label{fig.0}
\end{figure}

\subsection{Sample preparation}
\label{Sample}

Our working system is made of carbopol ETD~2050 dispersed in water and neutralized using sodium hydroxyde (NaOH). Carbopol ETD~2050 comprises homo- and copolymers of acry\-lic acid that are highly crosslinked with a polyalkenyl poly\-ether \cite{Roberts:2001}. Above a certain amount of carbopol in the dispersion, the system is made of an assembly of soft jammed swollen polymer ``particles,'' whose size ranges roughly from a few microns to 20~$\mu$m \cite{Ketz:1988,Kim:2003,Oppong:2006,Lee:2011}. Such a microstructure is known as a {\it microgel} \cite{Ketz:1988}.

Carbopol microgels were shown to be non-aging, non-thi\-xo\-tro\-pic simple YSF \cite{Piau:2007,Coussot:2009,Moller:2009a,Divoux:2010} and to exhibit good temperature stability \cite{Islam:2004}. Their steady-state flow curve is well described by the Herschel-Bulkley constitutive equation \cite{Piau:2007,Moller:2009a,Divoux:2010}. We also emphasize that the microgel macroscopic properties depend on the details of the preparation protocol. Indeed, the microgel preparation traditionally involves two steps: (i) the polymer is dispersed in water at pH$<7$ leading to a liquidlike suspension of carbopol aggregates and (ii) a neutralizing agent [most often NaOH or triethanolamine (TEA)] is added, which induces polymer swelling and microgel formation. During the second preparation step, the way the base is added (drop by drop or all at once) as well as the final value of the pH were reported to influence the rheological properties of the resulting microgels \cite{Curran:2002,Lee:2011}. Moreover, it was found that the stirring speed during the neutralization process also influences the particle size distribution and thus the properties of the microgel \cite{Baudonnet:2004}. Therefore, if one wishes to compare quantitative results obtained in different geometries, with different experimental protocols, or under different boundary conditions, the experiments must be performed on the same batch of carbopol so that the chemistry and the influence of the microgel preparation are not at stake. We shall pay attention to this issue throughout the present paper.

For our study, two kinds of samples are prepared: (i) traditional samples as described above, referred to as ``pure'' samples, and (ii) samples seeded with micronsized hollow glass spheres at 0.5~\%~w/w (Potters, Sphericel, mean diameter 6~$\mu$m, density 1.1), referred to as ``seeded'' samples. These hollow glass spheres act as acoustic contrast agents for ultrasonic spe\-ckle velocimetry (USV, see details below). We shall show that seeded samples exhibit very similar rheological properties as pure samples and that the addition of hollow glass spheres does not affect the properties of the stress overshoot.

The detailed preparation protocol for a {\it seeded} sample is as follows. We first suspend 0.5~\%~w/w of hollow glass spheres in ultrapure water; pH increases roughly from 7 to 8. Since carbopol is hydrosoluble only for pH$<7$, we add one or two drops of concentrated acid in order to make the pH drop to about 5. The glass sphere suspension is then heated at $50^{\circ}$C and the carbopol powder is carefully dispersed under magnetic stirring at 300~rpm for 40~min, at a weight fraction $C$ ranging from 0.5 to 3~\%~w/w. The mixture is then left at rest at room temperature for another 30~min, after which pH~$\simeq 3$. Finally, we neutralize the solution by adding NaOH (at a concentration of 10 mol.L$^{-1}$) drop by drop until pH~$=7.0\pm 0.5$ under vigorous manual stirring. This leads to a carbopol microgel which is finally centrifuged for 10~min at 2500~rpm to get rid of trapped bubbles. For a {\it pure} sample, the preparation protocol starts directly by adding the carbopol powder into a heated volume of ultrapure water and continues as explained above. In the following, we present results obtained on different batches of pure samples at carbopol concentrations $C=0.5$, 1, 2 and 3~\% w/w and seeded samples at a carbopol concentration $C=1$~\% w/w.  

\begin{figure*}[t!]\tt
\centering
\includegraphics[width=0.6\linewidth]{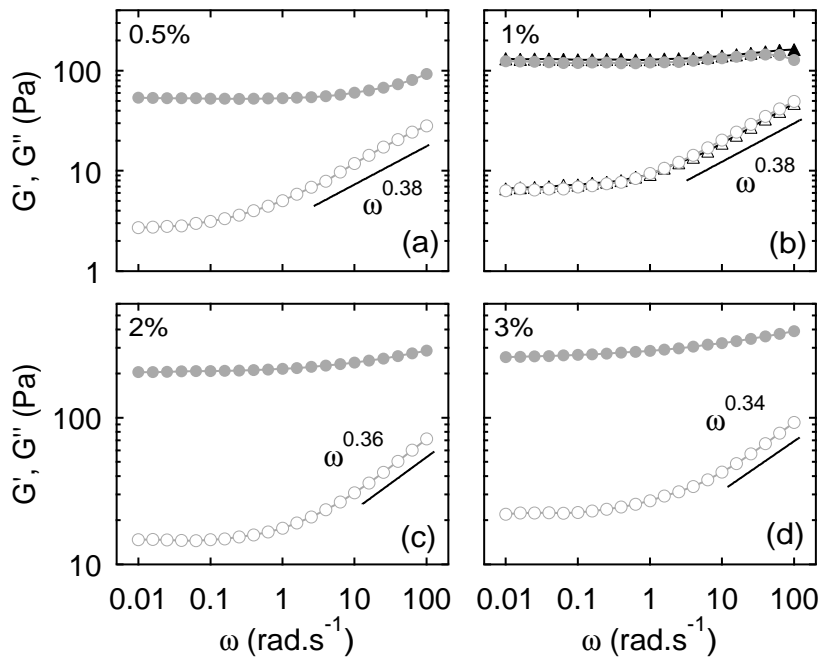}
\caption{Elastic modulus $G'$ ($\bullet$) and viscous modulus $G''$ ($\circ$) vs frequency $\omega$ for pure samples at (a) $C=0.5$, (b) 1, (c) 2 and (d) 3~\% w/w. In (b), the triangles correspond to a carbopol microgel at 1~\% w/w seeded with 0.5~\% w/w hollow glass spheres. The imposed strain amplitude is $\gamma = 1$~\%. For all concentrations, the elastic modulus increases very weakly with the frequency while the viscous modulus increases as a power law of the frequency for $\omega \gtrsim 1$~rad.s$^{-1}$. The whole data set was obtained in a plate-and-plate geometry of gap $e=1$~mm under rough boundary conditions (sand paper of roughness 46 $\mu$m).}
\label{fig.1}
\end{figure*}

\subsection{Experimental setups and protocol}
\label{Protocolexp}

\subsubsection{Rheological setup}.~Rheological measurements are performed using a stress-controlled rheometer (Anton Paar, MCR 301) in two different shearing geometries: a cylindrical Couette geometry (rotating inner cylinder radius 23.5 mm, gap width $e=1.1$~mm, and height 28 mm) and a plate-and-plate geometry (radius 21~mm, gap width $e=1$~mm). In both cases, sand paper was glued on both shearing surfaces to provide a roughness of 60~$\mu$m (resp. 46~$\mu$m) in the case of the Couette (resp. plate-and-plate) geometry. Such a roughness was chosen to be of the order of the microstructure size in order to minimize wall slip \cite{Meeker:2004a,Meeker:2004b}.

In Section \ref{bc} below, we shall explore the influence of boundary conditions by changing the surface roughness. In the Couette geometry, polished Plexiglas surfaces (rotating inner cylinder radius 24 mm, gap width $e=1$~mm, and height 28 mm), referred to as ``smooth'' Plexiglas walls, will be used to provide a surface roughness of $\delta\simeq 15$~nm as measured from atomic force microscopy (AFM). In the plate-and-plate geometry, a set of plates made of glass ($\delta\simeq 6$~nm from AFM), roughened Plexiglas ($\delta\simeq 1~\mu$m from AFM), and different glued sand papers will allow us to vary the surface roughness more finely. In all cases, the working temperature is 25$^\circ$C.

\subsubsection{Local velocity measurements}.~Velocity profiles across the gap of the Couette cell are recorded in seeded samples simultaneously to the global rheological data through ultrasonic speckle velocimetry (USV) as described by Manneville {\it et al.} \cite{Manneville:2004a}.

In short, USV relies on the analysis of ultrasonic speckle signals that result from the interferences of the backscattered echoes of successive incident pulses of central frequency 36~MHz generated by a high-frequency piezo-polymer transducer (Panametrics PI50-2) connected to a broadband pulser-receiver (Panametrics 5900PR with 200 MHz bandwidth). The speckle signals are sent to a high-speed digitizer (Acqiris DP235 with 500 MHz sampling frequency) and stored on a PC for post-processing using a cross-correlation algorithm that yields the local displacement from one pulse to another as a function of the radial position across the gap with a spatial resolution of 40~$\mu$m. After a calibration step using a Newtonian fluid, tangential velocity profiles are then obtained by averaging over 10 to 1000 successive cross-correlations depending on the desired temporal resolution. Full technical details about USV may be found in Ref.~\cite{Manneville:2004a}.

Here, the sample velocity field is measured at about 15~mm from the cell bottom. As already mentioned above, since pure carbopol microgels are transparent to ultrasound, we consider samples seeded with 0.5~\% w/w hollow glass spheres that provide acoustic contrast. The time needed to record a single velocity profile is inversely proportional to the applied shear rate. It is set to 1~s (10~s resp.) for the data shown in Fig.~\ref{fig.3} (in Fig.~\ref{fig.12} resp.). 

\subsubsection{Experimental protocol}.\label{Protocolexp3}~To ensure that the strain accumulated during loading into the cell has no influence, the sample is systematically presheared for 1 min at 1000~s$^{-1}$ and for 1 min at -1000~s$^{-1}$ before any measurement. We then check that a reproducible initial state is reached by measuring the viscoelastic moduli at $\omega=1$~Hz under a small oscillatory strain of amplitude $\gamma=1~\%$. Figure~\ref{fig.0} shows that $G'(t)$ first increases steeply for $t\lesssim 10$~s and then follows a slow logarithmic growth. Such a logarithmic recovery of $G'$ after preshear indicates a very slow consolidation of the microgel over time as observed in stabilized suspensions of silica particles \cite{Derec:2003} and polyelectrolyte microgels \cite{Borrega:2000}. In any case, $G'$ no longer varies significantly after 3 min. Therefore, in our experiments below, we first define the elastic modulus $G_0$ of the microgel prior to each experiment as $G_0=G'(\omega=1$~Hz$)$ measured 2 min after preshear. Next, the sample is left at rest for one more minute before starting the experiment so that the total {\it waiting time} between the end of preshear and the start of the actual measurement is $t_w=180$~s. This waiting time will be varied in Section~\ref{protocol} only, where we shall investigate the effect of $t_w$ on the stress overshoot phenomenon.

We also carefully checked that imposing $+\dot \gamma$ or $-\dot \gamma$ after the preshear yields exactly the same stress response. Thus, we do not see any dependence on the rotation direction during preshear contrary to the observations by Mahaut {\it et al.} on a carbopol 980 microgel \cite{Mahaut:2008} (see Fig.~4 of this reference). In our protocol, the total strain imposed to the microgel prior to each experiment $\gamma = 120,000$ is much larger than the one imposed by Mahaut {\it et al.}, $\gamma = 600$, which may have been insufficient to fully erase the loading history.

\begin{figure}[tb]\tt
\centering
\includegraphics[width=0.9\linewidth]{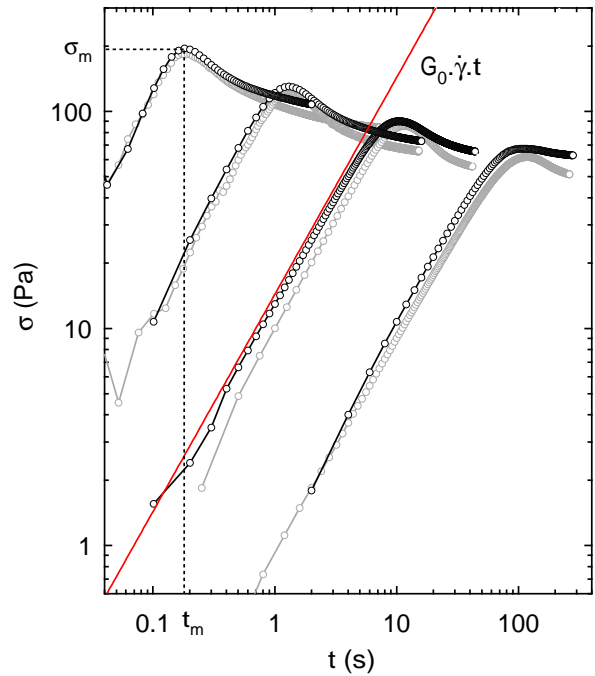}
\caption{Shear stress $\sigma$ versus time $t$ for various shear rates applied at $t=0$: $\dot \gamma =10, 1, 0.1$ and 0.01~s$^{-1}$ from left to right. Black (resp. grey) symbols correspond to a pure (resp. seeded) carbopol microgel at $C=1$~\% w/w. The stress first increases linearly with time. This corresponds to elastic deformation as confirmed by the red line showing $\sigma(t)=G_0\dot \gamma t$, for $\dot\gamma=0.1$~s$^{-1}$, where $G_0=143.5$~Pa is the elastic modulus of the pure microgel measured independently prior to shear start-up (see Section~\ref{Protocolexp3}). The stress reaches its maximum $\sigma_m$ at a time $t_m$ before slowly decreasing towards its steady-state value. Both data sets were obtained in a plate-and-plate geometry of gap $e=1$~mm under rough boundary conditions (sand paper of roughness 46 $\mu$m).}
\label{fig.2}
\end{figure}

\section{Results}

\subsection{Linear rheology}

\begin{figure*}[!t]\tt
\centering
\includegraphics[width=0.6\linewidth]{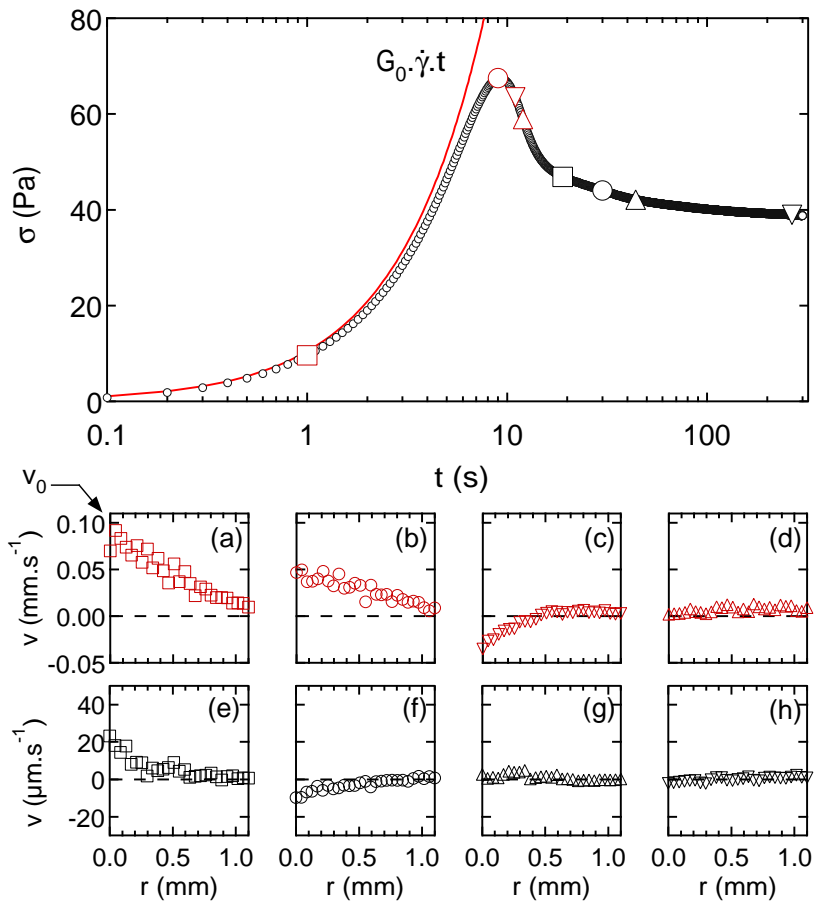}
\caption{Top: Shear stress $\sigma$ vs time $t$ for a shear rate $\dot \gamma = 0.1$~s$^{-1}$ applied at $t=0$. The red line shows $\sigma(t)=G_0\dot\gamma t$, where $G_0=104.3$~Pa is measured prior to shear start-up. Bottom: Velocity profiles $v(r)$ , where $r$ is the distance to the rotor, at different times [(letter), symbol, time (s)]: [(a), $\square$, 1]; [(b), $\circ$, 9]; [(c), $\triangledown$, 11]; [(d), $\vartriangle$, 12]; [(e), $\square$, 19]; [(f), $\circ$, 30]; [(g), $\vartriangle$, 44]; [(h), $\triangledown$, 266]. The rotor velocity $v_0$ is indicated by an arrow and corresponds to the upper bound of the vertical axis in (a)--(d). Note that the vertical scale in (e)--(h) is three times larger in order to emphasize small velocities. Experiments performed on a seeded 1~\% w/w carbopol microgel in a Couette cell of gap $e=1.1$~mm under rough boundary conditions (sand paper of roughness 60~$\mu$m).}
\label{fig.3}
\end{figure*}

In order to probe the viscoelastic properties of our samples, we first perform decreasing frequency sweeps under strain oscillations of small amplitude in the linear range ($\gamma = 1$~\%). As shown in Fig.~\ref{fig.1}, the elastic modulus $G'$ at low frequency is always more than 10 times larger than the loss modulus $G''$ for all concentrations. $G'$ is found to increase very weakly with the frequency $\omega$ whereas $G''$ presents two different regimes: $G''$ remains constant at frequencies lower than $\simeq 1$~rad.s$^{-1}$ and increases as a power law, $G''(\omega)\propto \omega^\xi$, for $\omega \gtrsim 1$~rad.s$^{-1}$. Table~\ref{tbl:linrheol} gathers the elastic modulus $\overline{G}_0$ averaged over $\omega=0.01$--1~rad.s$^{-1}$ as well as the values of the exponent $\xi$ for the different samples. $\overline{G}_0$ is observed to be roughly proportional to the carbopol concentration while $\xi$ slightly decreases with $C$. Note that all these results are in agreement with previous work on microgels made of carbopol 940 \cite{Benmouffok:2010} and 941 \cite{Ketz:1988}.

\begin{table}[h]
\small
\caption{Characteristic properties of our various carbopol microgels extracted from the data in Fig.~\ref{fig.1}. The elastic modulus $\overline{G}_0$ is computed as an average of $G'(\omega)$ over $\omega=0.01$--1~rad.s$^{-1}$ and is given together with the corresponding standard deviation. The exponent $\xi$ of the viscous modulus is extracted from the best power-law fit $G''(\omega) \propto \omega^{\xi}$ for $\omega \gtrsim 1$~Hz.}
  \label{tbl:linrheol}
  \begin{tabular*}{0.5\textwidth}{@{\extracolsep{\fill}}lllll}
    \hline
     $C$ (\% w/w) & $\overline{G}_0$ (Pa) & $\xi $ \\
    \hline
    0.5 &  53$\pm$0.1	& $0.38\pm 0.01$\\
    1 (pure sample)&  121$\pm$2 	& $0.38\pm 0.01$\\
    1 (seeded sample)&  130$\pm$1 	& $0.39\pm 0.01$\\
    2 &  210$\pm$4 	& $0.36\pm 0.01$\\
    3 &  270$\pm$10	& $0.34\pm 0.01$\\
    \hline
  \end{tabular*}
\end{table}

Furthermore, the influence of the hollow glass spheres is tested on the 1~\% w/w microgel in Fig.~\ref{fig.1}(b): both the elastic and the viscous moduli collapse on the same curves to within about 20~\% (see also Table~\ref{tbl:linrheol}). Such a dispersion is of the same order as that observed from batch to batch for 1~\% w/w microgels due to the sensitivity on the preparation protocol as explained in Section~\ref{Sample} above. We conclude that linear rheological properties are not significantly altered by the presence of the acoustic contrast agents used for USV.

The linear viscoelasticity shown in Fig.~\ref{fig.1} is strongly reminiscent of other simple YSF, such as wet foams \cite{Cohen-Addad:1998,Gopal:2003} and emulsions \cite{Mason:1995a,Mason:1995c,Hebraud:2000} which present the exact same trends: an elastic modulus that is roughly independent of the frequency and a power-law behaviour for $G''$ at high frequencies, with an exponent $\xi \simeq 0.5$. On the one hand, the fact that $G'$ remains constant and much larger than $G''$ is interpreted in both systems as the signature of a gel-like or jammed structure. On the other hand, the power-law increase of $G''$ is linked to collective motions and interpreted as the slip of weak planes: at high frequencies, the material is more likely to involve the slip of large regions than to deform under applied strain. An elastic approach under this assumptions indeed predicts the observed scaling law \cite{Liu:1996b}. Finally, the absence of any noticeable downturn of $G''$ towards linear behaviour at low frequencies points to very slow relaxation modes typical of soft glassy materials \cite{Mason:1995c,Sollich:1997}.

\subsection{Stress overshoot during start-up experiments: typical stress response and velocity profiles}
\label{s.velocityprofiles}

Start-up experiments go as follows: a constant shear rate $\dot \gamma$ is imposed at time $t=0$ and the shear stress response $\sigma(t)$  is monitored for at least three strain units. Figures~\ref{fig.2} and \ref{fig.3}(a) show $\sigma(t)$ measured on pure and seeded microgels at a carbopol concentration of 1~\% w/w in rough plate-and-plate and Couette geometries. These stress responses are typical of all our carbopol samples under rough boundary conditions. They are characterized by an {\it overshoot} in that $\sigma(t)$ first increases linearly with time before passing through a maximum and finally decreasing to a steady-state value. As shown by the red lines in Figs.~\ref{fig.2} and \ref{fig.3}(a), the initial linear growth of the stress is given by $\sigma(t)=G_0 \dot\gamma t$, where $G_0$ is the elastic modulus of the microgel. This first regime suggests that the microgel undergoes a purely elastic deformation at short times. At a time $t = t_m$, $\sigma(t)$ reaches a maximum value $\sigma_m$ which is sometimes referred to in the literature as the {\it dynamic} \cite{Varnik:2004,Xu:2006} or as the {\it static yield stress} \cite{Khan:1988} of the material. To avoid any confusion, we will simply call $\sigma_m$ the maximum shear stress. In the final relaxation stage, $\sigma(t)$ slowly decreases towards its steady-state value. In a previous work \cite{Divoux:2010}, we have shown that three strain units are far from sufficient to reach steady-state. In fact, following the short-time stress overshoot, the long-time relaxation involves the nucleation and growth of a transient shear band that progressively fluidizes the whole sample. The fluidization time $\tau_f$ follows a non-trivial scaling law with the applied shear rate, $\tau_f\sim\dot\gamma^{-\alpha}$ with $\alpha=2$--3, so that it can take more than $\gamma=10^4$ strain units to reach steady state at low shear rates (see Fig.~2 in Divoux {\it et al.} \cite{Divoux:2010} for a typical long-time stress relaxation). Here, we only focus on the stress overshoot but one should keep in mind that the state reached after a strain $\gamma\sim 3$ does {\it not} correspond to steady state.
 
\begin{figure*}[t]\tt
\centering
\includegraphics[width=0.5\linewidth]{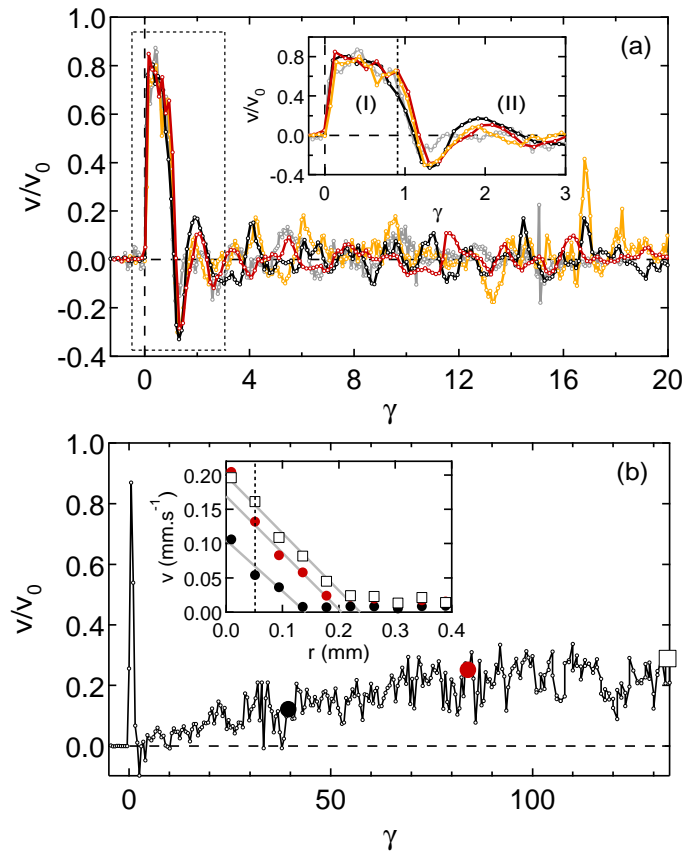}
\caption{(a) Velocity $v(r=50~\mu$m$,t)$ normalized by the rotor velocity $v_0$ and plotted against the strain $\gamma=\dot\gamma t$ for various shear rates $\dot \gamma = 5.10^{-2}$ (grey), $7.10^{-2}$ (orange), 0.1 (black) and 0.13~s$^{-1}$ (red) applied at $t=0$. $v$ is averaged over $\pm 50~\mu$m around the mean position $r=50~\mu$m from the rotor. Inset: horizontal zoom over the first three strain units which emphasizes the elastic deformation, failure, and recoil of the microgel. For $\gamma < 0.9$, the velocity slowly decreases and the deformation is elasto-plastic [regime (I)]. For $\gamma > 0.9$, the velocity rapidly drops, becomes negative, and undergoes damped oscillations [region (II)]. (b) Same as (a) for $\dot \gamma = 0.5$~s$^{-1}$. Inset: velocity profiles $v(r)$ for $r<400~\mu$m and $t=79$ ($\bullet$), 168 (red dots), and 267~s ($\square$) showing the nucleation and growth of a shear band at the rotor. The vertical dotted line indicates the position $r=50~\mu$m at which the velocity plotted in the main figure is measured. Same experimental conditions as in Fig.~\ref{fig.3}.}
\label{fig.4}
\end{figure*}

In order to get better insight on the bulk dynamics during the stress overshoot, we measure simultaneously the stress response and velocity profiles using USV on a seeded carbopol microgel at 1~\% w/w in a rough Couette geometry of gap $e=1.1$~mm. As shown in Fig.~\ref{fig.3} for $\dot \gamma = 0.1$~s$^{-1}$, the velocity profile is linear during the initial growth of the stress [Fig.~\ref{fig.3}(a)]. Up to experimental uncertainty, the sample velocity reaches the rotor velocity at $r=0$. This means that the whole sample is {\it homogeneously} strained without significant wall slip. Around the stress maximum, the microgel is seen to {\it break} at the rotor [Fig.~\ref{fig.3}(b)], which leads to a large {\it elastic recoil} [Fig.~\ref{fig.3}(c)] and several {\it damped oscillations} of the bulk material [Fig.~\ref{fig.3}(d)-(f)] at the start of the stress relaxation phase. This observation allows us to call the time $t_m$ at which the stress maximum is reached the {\it failure time} and the corresponding strain $\gamma_m = \dot \gamma  t_m$, the {\it failure strain}. The microgel then enters a regime of {\it total wall slip} [Fig.~\ref{fig.3}(g)-(h)]: the local shear rate effectively felt by the microgel is vanishingly small and the velocity within the sample fluctuates around 0. In other words, the shear rate applied by the rheometer is fully absorbed at the rotor by a lubrication layer whose size is smaller than the USV spatial resolution of 40~$\mu$m. As recalled above, total wall slip is {\it not} the steady state. As shown in the inset of Fig.~\ref{fig.4}(b), it rather gives way to a transient shear band that nucleates at the rotor from the lubrication layer and slowly grows as the strain increases \cite{Divoux:2010}.

Figure~\ref{fig.4} further highlights our temporally-resolved velocity measurements by focusing on the velocity measured at $r=50~\mu$m from the rotor as a function of the strain $\gamma = \dot \gamma t$ for various applied shear rates. Comparing Fig.~\ref{fig.4}(a) to Fig.~\ref{fig.3}(top), we observe that, at small strains, the velocity starts to decrease before the stress reaches its maximum $\sigma_m$. Indeed, $\sigma_m$ is reached for a strain $\gamma_m\simeq 0.9$ while $v(r=50~\mu$m$)$ decreases from about $0.8\,v_0$ to $0.5\,v_0$ when $\gamma$ increases from 0.1 to 0.9. Thus the initial increase of $\sigma (t)$ does not correspond to a purely elastic deformation and {\it plasticity} has to occur, which prepares the failure of the microgel at the rotor. This is confirmed by a closer inspection of the stress response: it can be seen on Fig.~\ref{fig.3}(top) that $\sigma(t)$ nicely coincides with $G_0 \dot\gamma t$ only for $t<1$~s (i.e. $\gamma<0.1$) and that $\sigma(t)$ lies slightly but significantly below the purely elastic response at larger strains. We conclude that the regime that precedes the microgel failure is {\it elasto-plastic} rather than purely elastic. Furthermore, the strong acceleration of the microgel around $\gamma = 0.9$ [see the fast increase of the slope in $v(t)$ in the inset of Fig.~\ref{fig.4}(a)] advocates for a scenario where the failure of the microgel at the wall originates from the accumulation of a large enough amount of plastic events. Note that under imposed stress and rough boundary conditions, we have shown that the shear rate follows the Andrade's creep scaling law, $\dot \gamma (t) \propto t^{-2/3}$, a behaviour which is also characteristic of plastic deformation \cite{Divoux:2011pp}. Figure~\ref{fig.4}(b) shows that for larger strains, the velocity close to the rotor slowly increases with time. This increase actually corresponds to the long-time nucleation and growth of a shear band already reported elsewhere \cite{Divoux:2010}. Although not under scrutiny in the present paper, the early stages of this transient shear-banding regime are shown in the inset of Fig.~\ref{fig.4}(b) through velocity profiles close to the rotor recorded at $\dot \gamma = 0.5$~s$^{-1}$ for strains up to $\gamma=130$.

\begin{figure}[!t]\tt
\centering
\includegraphics[width=0.9\columnwidth]{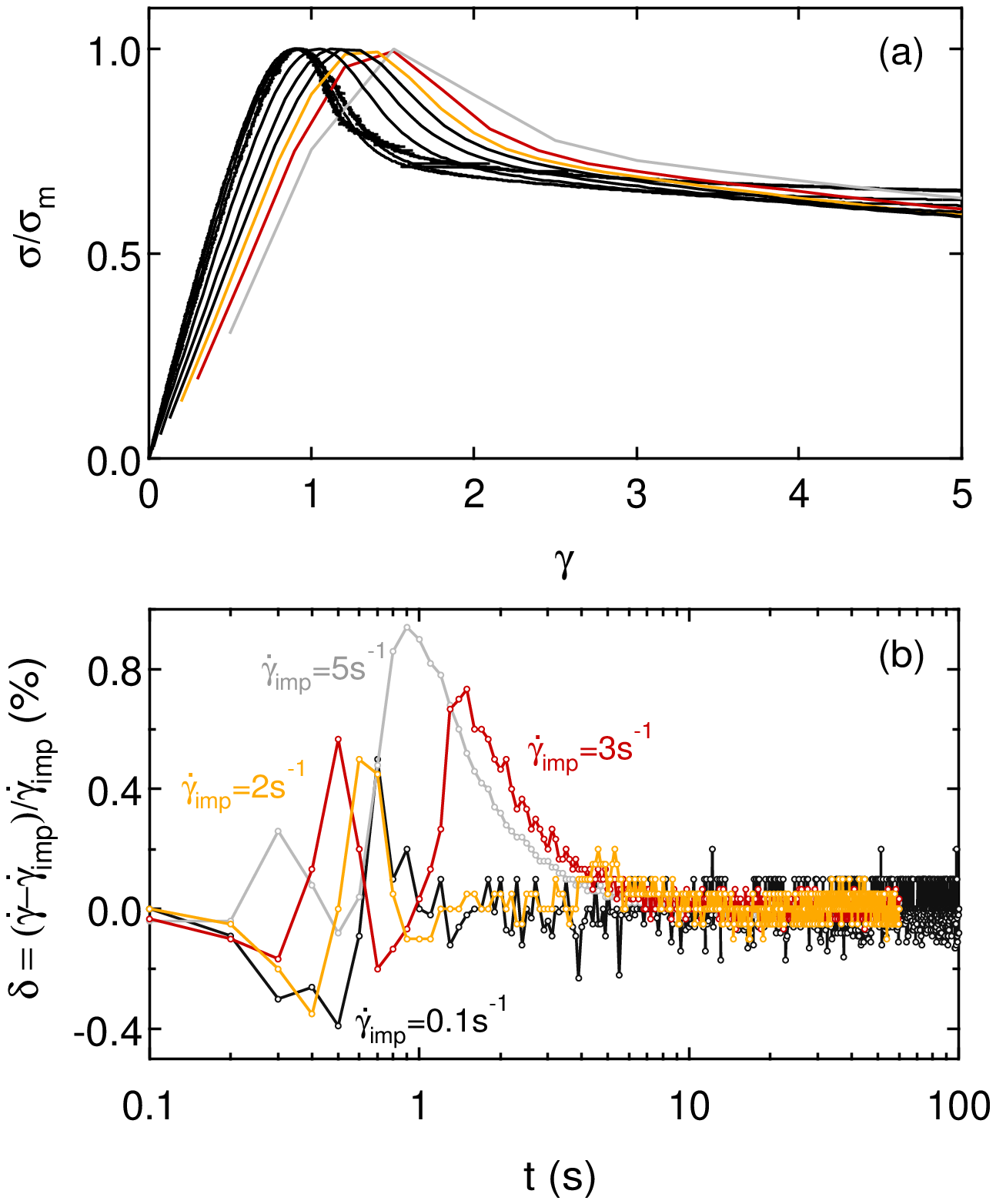}
\caption{(a) Normalized shear stress $\sigma/\sigma_m$ vs strain $\gamma$ for various applied shear rates: $\dot \gamma=5.10^{-4}$, $5.10^{-3}$, $5.10^{-2}$, 0.1, 0.3, 0.7, 1.3 (black lines), 2 (orange line), 3 (red line), and 5~s$^{-1}$ (grey line). For $\dot \gamma \lesssim 0.05$~s$^{-1}$, all the normalized shear stresses collapse on a single curve. For $\dot \gamma \gtrsim 0.05$~s$^{-1}$, the curves are shifted towards larger strains. (b) Relative difference $\delta $ between the commanded shear rate $\dot \gamma_{\rm imp}$ and the shear rate $\dot \gamma$ effectively applied and measured by the rheometer vs time $t$ for $\dot\gamma_{\rm imp}=0.1$ (black line), 2 (orange line), 3 (red line), and 5~s$^{-1}$ (grey line). The feedback loop of the rheometer allows for a good control of the shear rate since $\delta$ remains always smaller than 1~\% even at the highest imposed shear rates. Same experimental conditions as in Fig.~\ref{fig.3}.}
\label{fig.5}
\end{figure}

\subsection{Influence of the shear rate}

The above start-up experiments on a seeded carbopol microgel at 1~\% w/w in a rough Couette cell were repeated for various applied shear rates $\dot\gamma$ ranging from 10$^{-4}$ to 10~s$^{-1}$. The corresponding normalized stress responses $\sigma(t)/\sigma_m$ are shown as a function of the strain $\gamma$ in Fig.~\ref{fig.5}(a). While the stress responses all coincide for $\dot \gamma \lesssim 0.05$~s$^{-1}$, the failure strain clearly shifts to values larger than 1 for larger shear rates. Since at high shear rates large overshoots are reached within a rather short time (see, e.g. $\dot \gamma =10$~s$^{-1}$ in Fig.~\ref{fig.2}), it is important to check that the feedback loop of our stress-controlled rheometer ensures that the desired shear rate is effectively reached without significant perturbation due to the stress overshoot. Such a test is performed in Fig.~\ref{fig.5}(b) by considering the relative difference $\delta(t)$ between the actual shear rate $\dot\gamma(t)$ imposed to the sample by the rheometer and the commanded shear rate $\dot \gamma_{\rm imp}$. In all cases, $\delta(t)$ remains smaller than 1~\% so that we can exclude any artifact due to using a stress-controlled rheometer in the shear-rate controlled mode.

\begin{figure}[!t]\tt
\centering
\includegraphics[width=0.9\columnwidth]{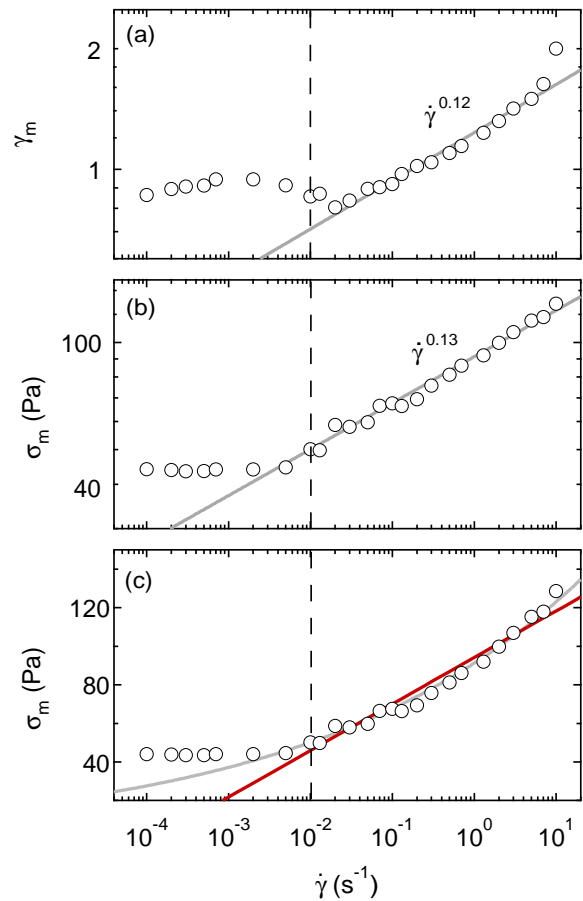}
\caption{(a) Failure strain $\gamma_m$ vs applied shear rate $\dot \gamma$ in logarithmic scales. The grey line is the best power-law fit, $\gamma_m = A\dot \gamma^\mu$ for $\dot\gamma>0.01$~s$^{-1}$, which yields $A = 1.23\pm 0.01$ and $\mu=0.12\pm 0.01$. (b) Maximum shear stress $\sigma_m$ vs applied shear rate $\dot \gamma$ in logarithmic scales. The grey line is the best power-law fit, $\sigma_m = A\dot \gamma^\nu$ for $\dot\gamma>0.01$~s$^{-1}$, which yields $A = 91.2\pm 0.7$ and $\nu=0.13\pm 0.01$. (c)~Same as (b) plotted in semilogarithmic scales. The previous power-law fit (grey line) provides a better description than the best logarithmic fit over $\dot \gamma>0.01$~s$^{-1}$ (straight red line). The vertical dashed lines indicate the critical shear rate $\dot\gamma_w\simeq 0.01$~s$^{-1}$ that separates the power-law regime from the low-shear regime where both $\gamma_m$ and $\sigma_m$ are independent of the shear rate. Same experimental conditions as in Fig.~\ref{fig.3}.}
\label{fig.6}
\end{figure}

The evolutions of the failure strain $\gamma_m$ and the maximum shear stress $\sigma_m$ with $\dot\gamma$ are reported in Fig.~\ref{fig.6}(a) and (b) respectively. Both quantities present two different regimes when plotted in logarithmic scales: for $\dot \gamma < \dot \gamma_w \simeq 10^{-2}$~s$^{-1}$, $\gamma_m$ and $\sigma_m$ remain roughly constant and independent of the applied shear rate whereas for $\dot \gamma > \dot \gamma_w$, they increases as weak power laws of the shear rate with similar exponents (0.12 and 0.13 respectively). It is checked in Fig.~\ref{fig.6}(c) that a logarithmic growth of $\sigma_m$ with $\dot\gamma$ does not provide as good a description of the experimental data as a power law: when plotted using semilogarithmic scales, the $\sigma_m$ vs $\dot\gamma$ data show a clear upward curvature and the linear fit over $\dot \gamma > \dot \gamma_w$ (shown in red) is not as accurate as the power-law fit (shown in grey). The reason for this discussion is that the overshoot data reported for bidisperse Lennard-Jones mixtures have been analyzed in terms of a logarithmic law based on the Ree-Eyring's viscosity theory \cite{Varnik:2004,Rottler:2005}, $\sigma_m=\sigma_0+k_B T/v^\star \ln(\dot\gamma/\nu_0)$, where $\sigma_0$ is some constant shear stress, $v^\star$ is the volume of a region involved in an elementary shear motion (called ``hopping'' motion), and $\nu_0$ is the attempt frequency of hopping. On the other hand, both a fluidity model \cite{Derec:2003} and Brownian dynamics simulations of particle gel \cite{Whittle:1997} have reported power-law behaviours of $\sigma_m$ vs $\dot\gamma$ with an exponent $\nu\simeq 0.5$. Experiments on stabilized suspensions of silica particles \cite{Derec:2003} and very recent experiments on attractive colloids \cite{Koumakis:2011} have also unveiled power laws with exponents $\nu\simeq 0.27$ and $\nu\simeq 0.5$ respectively. Here we report even smaller values of $\nu$ down to 0.13. As discussed below in Section~\ref{discuss}, the reason for such a variety of exponents remains unclear and stands out as an open question. The existence of the critical shear rate $\dot \gamma_w$ separating the two different regimes is addressed in detail in the following section.

\begin{figure}[!t]\tt
\centering
\includegraphics[width=0.9\columnwidth]{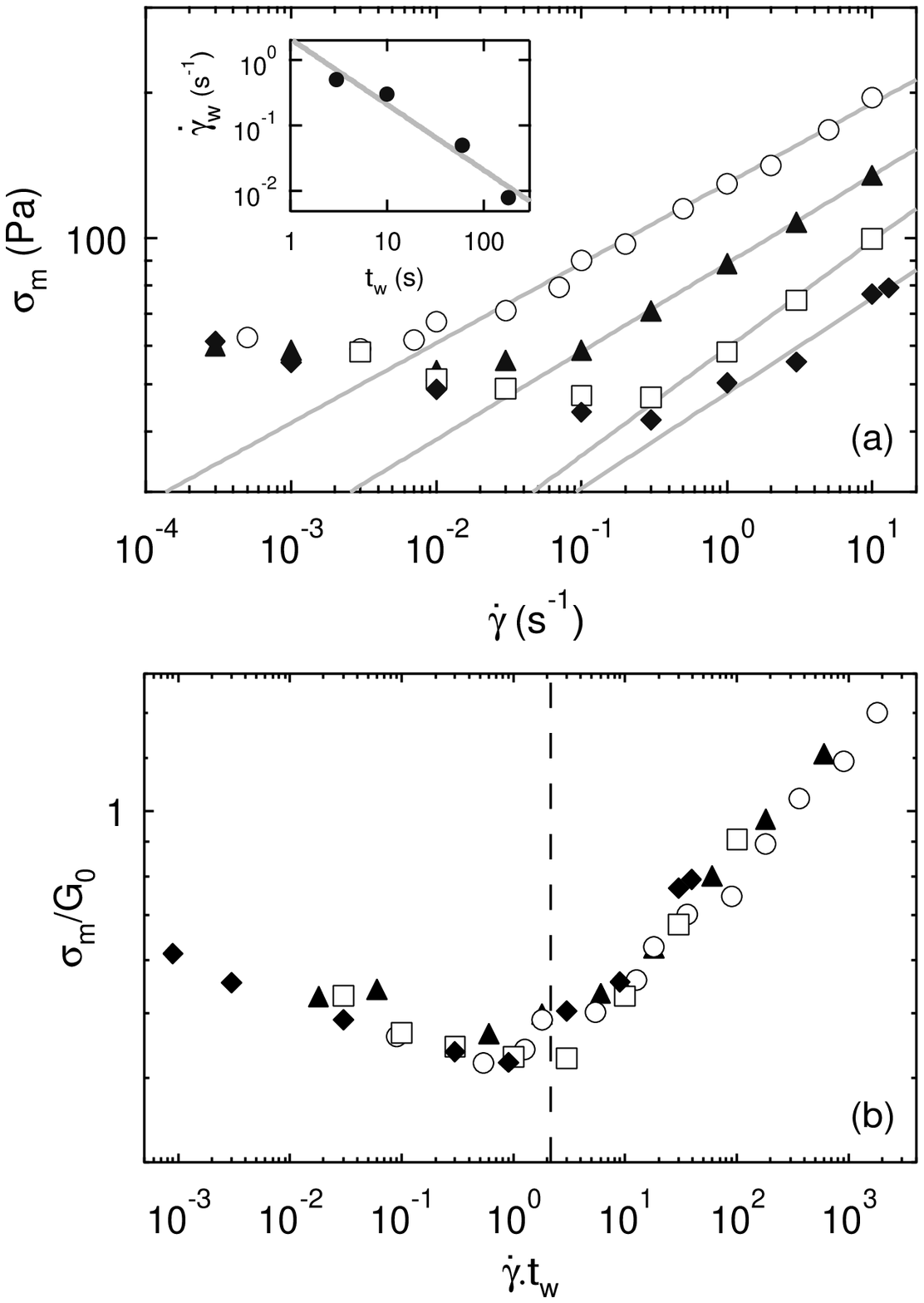}
\caption{(a) Maximum shear stress $\sigma_m$ vs applied shear rate $\dot \gamma$ for various waiting times $t_w$ [symbol, $t_w$ (s)]: ($\blacklozenge$, 3); ($\square $, 10); ($\blacktriangle$, 60); ($\circ$, 180). The best power-law fits, $\sigma_m = A \dot \gamma ^\nu$, obtained for $\dot \gamma > \dot \gamma_w$ are shown as grey lines and the corresponding fit parameters are gathered in Table~\ref{tbl:tw}. Inset: critical shear rate $\dot \gamma_w$ vs waiting time $t_w$. The grey line is the best linear fit in logarithmic scales: $\dot \gamma_w = 2.1/t_w$. (b) Rescaled data $\sigma_m/G_0$ vs $\dot\gamma t_w$. The vertical dashed line corresponds to $\gamma_w=\dot\gamma t_w=2.1$. Experiments performed on a pure 1~\% w/w carbopol microgel in a plate-and-plate geometry of gap $e=1$~mm under rough boundary conditions (sand paper of roughness 46 $\mu$m).}
\label{fig.7}
\end{figure}

\subsection{Influence of the protocol}
\label{protocol}

In this section, we investigate the influence of the experimental protocol on the stress overshoot and, more specifically, the influence of the waiting time $t_w$ between the preshear and the start of the experiment (see Section~\ref{Protocolexp}). To this aim, we performed start-up experiments for various applied shear rates and for different waiting times. Figure~\ref{fig.7}(a) gathers the $\sigma_m$ vs $\dot\gamma$ data obtained on a pure 1~\% w/w sample in a rough plate-and-plate geometry for $t_w=3$, 10, 60, and 180~s. As already observed in Fig.~\ref{fig.6}(b), one can define a critical shear rate $\dot \gamma_w$ below which the maximum shear stress is constant (or decreases very slowly with $\dot\gamma$ for the smallest waiting times, a feature that will be further discussed in Section~\ref{discuss}) and above which it increases as a power law of the shear rate. Table~\ref{tbl:tw} gathers the parameters of the best power-law fits obtained on the data of Fig.~\ref{fig.7}(a). In the power-law regime, $\sigma_m$ increases with the waiting time while for $\dot\gamma<\dot\gamma_w$ the data are undistinguishable. In view of the error bars, the overall slight decrease of the exponent from 0.2--0.22 at small $t_w$ to 0.16 for $t_w=180$~s is believed to be insignificant. However, the critical shear rate $\dot \gamma_w$ clearly decreases as $t_w$ increases: as shown in the inset of Fig.~\ref{fig.7}(a), one has $\dot \gamma_w =2.1/t_w$. We conclude that the value of $\dot \gamma_w$ is fixed by the waiting time $t_w$ between the preshear and the start of the experiment. Incidentally, it is also worth noticing that the $\sigma_m$ data shown here for $t_w=180$~s and a plate-and-plate geometry is quantitatively close to the $\sigma_m$ data measured in a Couette cell and shown in Fig.~\ref{fig.6}(b), so that the geometry has little influence on the stress overshoot phenomenon.
 
\begin{figure}[!t]\tt
\centering
\includegraphics[width=0.9\columnwidth]{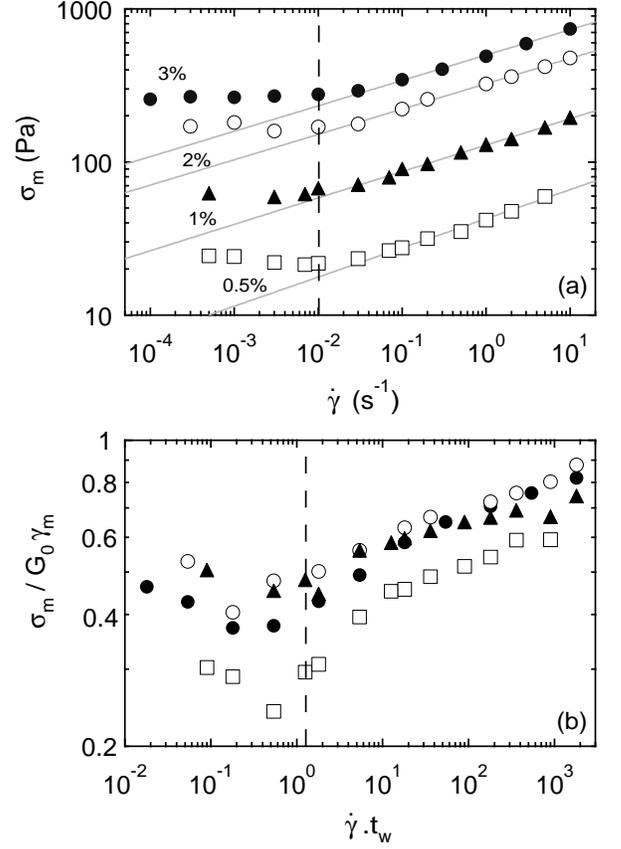}
\caption{(a) Maximum shear stress $\sigma_m$ vs applied shear rate $\dot \gamma$ for various carbopol weight fractions $C$ (symbol, $C$ \% w/w): ($\square$, 0.5); ($\blacktriangle$, 1); ($\circ $, 2); ($\bullet$, 3). The best power-law fits, $\sigma_m = A \dot \gamma ^\nu$, obtained for $\dot \gamma > \dot \gamma_w$ are shown as grey lines and the corresponding fit parameters are gathered in Table~\ref{tbl:gelconcentration}. The vertical dashed line indicates $\dot\gamma_w \simeq10^{-2}$~s$^{-1}$. (b) Rescaled data $\sigma_m/(G_0 \gamma_m)$ vs $\dot\gamma t_w$. The vertical dashed line corresponds to $\gamma_w=\dot\gamma t_w=2.1$. Experiments performed on pure carbopol microgels in a plate-and-plate geometry of gap $e=1$~mm under rough boundary conditions (sand paper of roughness 46 $\mu$m).}
\label{fig.8}
\end{figure} 
 
\begin{table}[h!]
\small
\caption{Parameters for various waiting times extracted from the data of Fig.~\ref{fig.7}(a): average $\overline{G}_0$ of the elastic moduli $G_0=G'(\omega=1$~Hz$)$ measured prior to each experiment for $t_w\ge 60$~s (together with the corresponding standard deviation), the critical shear rate $\dot\gamma_w$, the prefactor $A$, and the exponent $\nu$ of the best power-law fit $\sigma_m=A\dot \gamma^{\nu}$ for $\geq \dot \gamma_w$.}
\label{tbl:tw}
  \begin{tabular*}{0.5\textwidth}{@{\extracolsep{\fill}}llllll}
    \hline
     $t_w$ (s) & $\overline{G}_0$ (Pa) & $\dot\gamma_w$ (s$^{-1}$) & $A$ (Pa.s$^{\nu}$) & $\nu$ \\
    \hline
    3    & 100$^{\star}$  &   0.5	 & $48\pm 2$ 	& $0.20 \pm 0.03$\\
    10   & 110$^{\star}$  & 0.3	 & $59\pm 1$ 	& $0.22 \pm 0.01$\\
    60   & $112 \pm 2$  & 0.05	 & $89\pm 1$ 	& $0.18 \pm 0.01$\\
    180  & $141 \pm 5$  & 0.008 & $130\pm 2$ 	& $0.16 \pm 0.01$\\
    \hline
    \hline
  \end{tabular*}
\end{table}

The similar power laws obtained for the various waiting times and the observed dependence of $\dot \gamma_w$ vs $t_w$ suggest that plotting the data as a function of $\dot \gamma t_w$ should lead to a universal behaviour. Such a rescaling is shown to collapse all the $\sigma_m$ data as long as one considers the maximum shear stress normalized by the elastic modulus $G_0$ of the microgel [see Fig.~\ref{fig.7}(b)]. For $t_w\ge 60$~s, the values of $G_0$ used in Fig.~\ref{fig.7}(b) are those obtained from small amplitude oscillatory shear at 1~Hz prior to each experiment. The values reported in Table~\ref{tbl:tw} for $t_w\le 10$~s and indicated by an asterisk were extrapolated from independent time-resolved measurements similar to those of Fig.~\ref{fig.0}.

The main result of this section is the scaling of the critical shear rate $\dot \gamma_w$ as $1/t_w$. A very similar phenomenology of the influence of $t_w$ on the maximum shear stress has already been reported for binary Lennard-Jones glasses \cite{Varnik:2004}. Such crossover was associated with shearing the system faster than its structural relaxation, the waiting time $t_w$ being comparable to the time for a particle to escape from local cages.

\subsection{Influence of the microgel concentration}

The stress overshoot phenomenon depicted above for 1~\% w/w carbopol microgels is very robust to a concentration change. Here, we vary the carbopol weight fraction from 0.5 to 3~\% w/w for a given waiting time $t_w=180$~s. As seen in Fig.~\ref{fig.8}(a), the behaviour of the stress maximum $\sigma_m$ remains unchanged: increasing the microgel concentration shifts the value of the stress overshoot toward higher values but the shape of the curve $\sigma_m(\dot \gamma)$ remains the same. In particular, as reported in Table~\ref{tbl:gelconcentration}, the exponent of the power law does not depend significantly on the concentration similarly to what is observed for attractive colloids \cite{Koumakis:2011}, and the shear rate $\dot\gamma_w$ that characterizes the crossover to power-law behaviour is also independent of $C$.

\begin{table}[b]
\small
\caption{Parameters for various carbopol concentrations extracted from the data of Fig.~\ref{fig.8}(a): average $\overline{G}_0$ of the elastic moduli $G_0=G'(\omega=1$~Hz$)$ measured prior to each experiment and average $\overline{\gamma}_m$ of the failure strains taken over all the shear rates (together with the corresponding standard deviations), the prefactor $A$ and the exponent $\nu$ of the best power-law fit $\sigma_m=A\dot \gamma^{\nu}$ for $\geq \dot \gamma_w$.}
\label{tbl:gelconcentration}
\begin{tabular*}{0.5\textwidth}{@{\extracolsep{\fill}}lllll}
    \hline
     $C$ (\% w/w) & $\overline{G}_0$ (Pa) & $\overline{\gamma}_m$ & $A$ (Pa.s$^{\nu}$) & $\nu$ \\
    \hline
    0.5 & $71 \pm 2$    &	$1.04\pm 0.18$ & $43\pm 1$ 	& $0.19 \pm 0.01$\\
    1   & $141 \pm 5$ &  $1.19\pm 0.33$   & $129\pm 2$ 	& $0.17 \pm 0.01$\\
    2   & $285 \pm 10$  &  $1.43\pm 0.29$  & $324\pm 5$ 	& $0.16 \pm 0.01$\\
    3   & $410 \pm 9$  &   $1.65\pm 0.23$  & $500\pm 8$ 	& $0.16 \pm 0.01$\\
    \hline
  \end{tabular*}
\end{table}

We found that the data in Fig.~\ref{fig.8}(a) could be collapsed onto a single curve for $C\ge 1$~\% w/w by normalizing $\sigma_m$ by $G_0\gamma_m$ [see Fig.~\ref{fig.8}(b)]. Note that if we use $G_0$ to normalize the stress maximum as in Fig.~\ref{fig.7}(a) the collapse is not as good. The fact that the data set for $C=0.5$~\% w/w lies slightly below the other curves in Fig.~\ref{fig.8}(b) also hints at a more subtle influence of the concentration than a simple linear scaling with the elastic properties of the microgel. A more thorough study including local velocity measurements on seeded samples as well as microscopic visualization of the microgel deformation at various concentrations will be undertaken to clarify this point.

\begin{figure}[!t]\tt
\centering
\includegraphics[width=0.9\columnwidth]{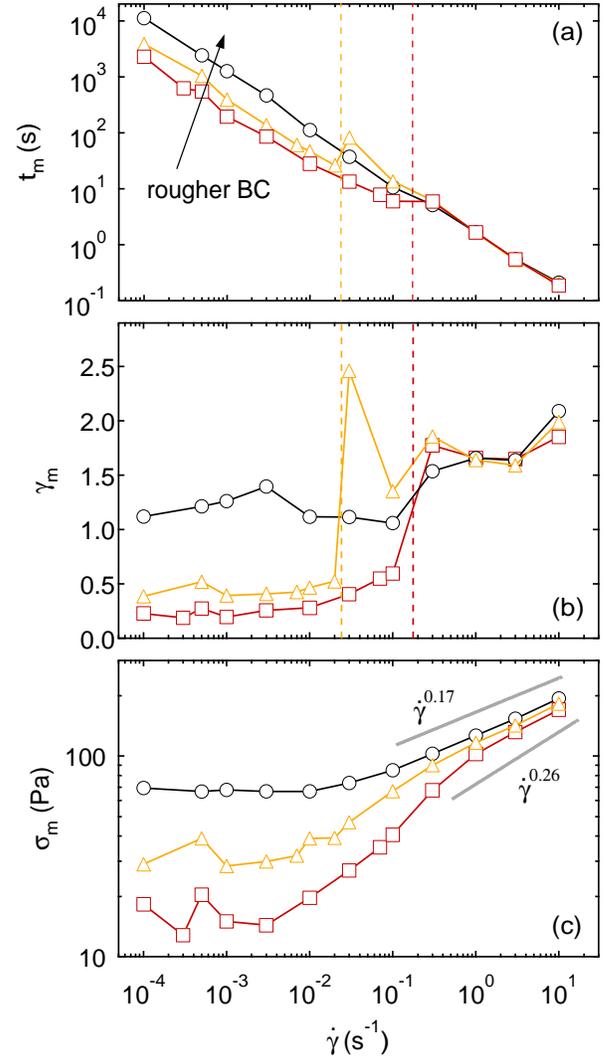}
\caption{(a) Failure time $t_m$, (b) failure strain $\gamma_m$, and (c) maximum shear stress $\sigma_m$ vs applied shear rate for various boundary conditions: smooth glass plates (red squares), roughened plexiglas (orange triangles), and glued sand paper (black circles). The vertical dotted lines in (a) and (b) indicate $\dot\gamma_s \simeq 0.02$ and 0.2~s$^{-1}$. Experiments performed on a pure 1~\% w/w carbopol microgel in plate-and-plate cells of gap $e=1$~mm with different surface roughnesses.}
\label{fig.9}
\end{figure}

\begin{figure}[!t]\tt
\centering
\includegraphics[width=0.9\columnwidth]{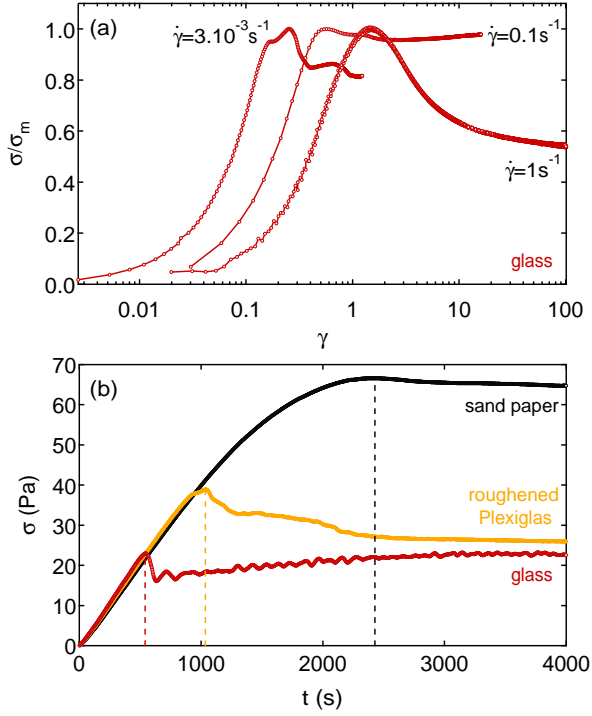}
\caption{(a) Normalized shear stress $\sigma/\sigma_m$ vs strain $\gamma$ for various applied shear rates under smooth boundary conditions: $\dot \gamma=3.10^{-3}$, 0.1, and 1~s$^{-1}$ from left to right. (b)~Shear stress $\sigma$ vs time $t$ measured at a given shear rate $\dot\gamma= 5.10^{-4}$~s$^{-1}$ for various boundary conditions: smooth glass plates (red), roughened Plexiglas (orange), and glued sand paper (black). Same experimental conditions as in Fig.~\ref{fig.9}.}
\label{fig.10}
\end{figure}

\subsection{Influence of the boundary conditions}
\label{bc}

Slip effects are well known in the soft matter literature to have a strong impact on rheological measurements \cite{Barnes:1995,Buscall:2010}. Until now, we have focused on rough boundary conditions obtained by gluing sand paper directly to the shearing tools. The large roughness of the walls ($\delta=46$ or 64~$\mu$m) was supposed to suppress wall slip. Yet, the local velocity measurements of Section~\ref{s.velocityprofiles} have revealed a scenario in which the stress overshoot geometry corresponds to failure of the microgel at the moving wall followed by total wall slip. In this section, we address the issue of whether or not the stress overshoot phenomenon is sensitive to boundary conditions. We first present and analyze global stress overshoot data obtained in three plate-and-plate cells of three different roughnesses and then investigate in detail the local behaviour of a seeded 1~\% w/w carbopol microgel in a smooth Plexiglas Couette cell.

\subsubsection{Influence of the wall roughness on global rheology.}
\begin{figure}[!t]\tt
\centering
\includegraphics[width=0.9\columnwidth]{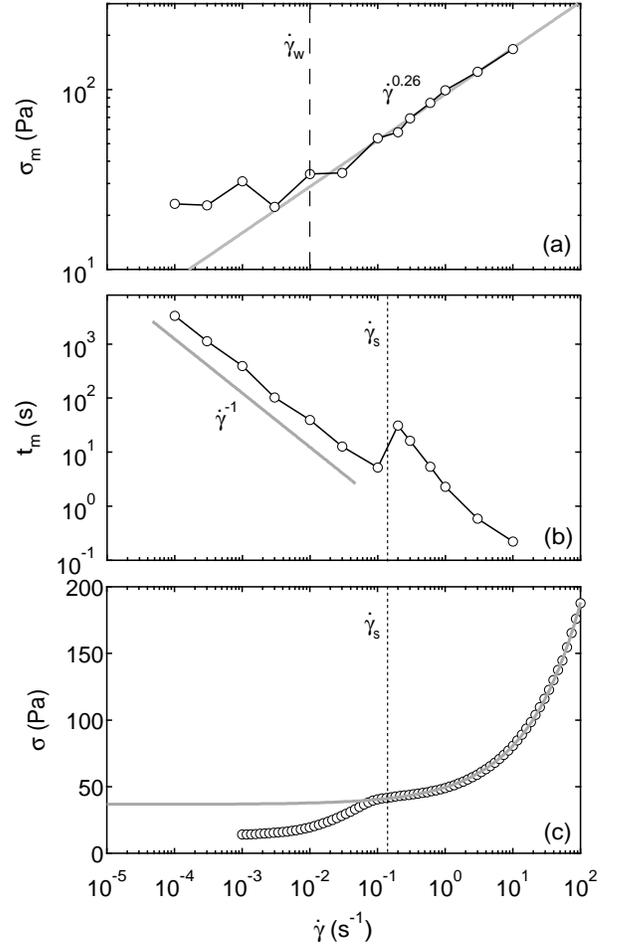}
\caption{(a) Maximum shear stress $\sigma_m$ vs applied shear rate. The best power-law fit for $\dot \gamma >  0.01$~s$^{-1}$ yields $\sigma_m = (94.1\pm 1.3).\dot \gamma^{(0.26\pm 0.01)}$ (grey line). (b)~Failure time $t_m$ vs applied shear rate $\dot \gamma$. The grey line has a slope -1. The vertical dotted line indicates $\dot\gamma_s \simeq 0.2$~s$^{-1}$. (c)~Steady-state flow curve, shear stress $\sigma$ vs shear rate $\dot \gamma$, obtained by decreasing $\dot \gamma$ from 100 to $10^{-3}$~s$^{-1}$ with a waiting time of 30~s per point. The grey line is the best fit by the Herschel-Bulkley (HB) model for $\dot\gamma>\dot\gamma_s \simeq 0.2$~s$^{-1}$: $\sigma = \sigma_c + A\dot \gamma^n$, with $\sigma_c = 36.7$ Pa, $n=0.54$, and $A=12.8$ Pa.s$^{-n}$. The deviation from the HB model for $\dot \gamma < \dot\gamma_s$ is the signature of wall slip at low shear rates. Experiments performed on a seeded 1~\% w/w carbopol microgel in a Couette cell of gap $e=1$~mm under smooth boundary conditions (polished Plexiglas of roughness 15~nm).}
\label{fig.11}
\end{figure}

~The overshoot phenomenon persists as the wall roughness is varied from rough to smooth so that one can still infer the global characteristics $t_m$, $\gamma_m$, and $\sigma_m$ of the overshoot as presented so far. Figure~\ref{fig.9} gathers such an analysis for overshoots recorded in plate-and-plate cells of gap 1~mm with three different wall roughnesses: $\delta=6$~nm obtained with glass plates, $\delta=1~\mu$m obtained with roughened Plexiglas, and $\delta=46~\mu$m obtained with sand paper. Even though the general trends reported above for rough boundary conditions are conserved, these data reveal significant effects of the wall roughness. At high shear rates, all data sets strikingly converge toward the case of rough boundary conditions. On the other hand, at low shear rates, smoother boundaries lead to much shorter failure times, and correspondingly to much smaller failure strains and maximum stresses. The shear rate at which the smoother cases are seen to roughly coincide with the rough case is noted $\dot\gamma_s$ and is found to be $\dot\gamma_s\simeq 0.02$~s$^{-1}$ for $\delta=1~\mu$m and $\dot\gamma_s\simeq 0.2$~s$^{-1}$ for $\delta=6~$nm [see dashed lines in Fig.~\ref{fig.9}(a) and (b)]. Moreover, although the $t_m$ and $\gamma_m$ data for smooth and rough walls indeed become undistinguishable up to experimental uncertainty for $\dot\gamma>\dot\gamma_s$, this is not truly the case for the stress maximum. $\sigma_m$ rather depends slightly on the boundary conditions even at the highest shear rates [see Fig.~\ref{fig.9}(c)]. Still, from the present data, it is not clear whether this dependence arises from that of the exponent of the power law that characterizes $\sigma_m$ or from its prefactor. The rather large difference between the exponents 0.17 and 0.26 of the power laws, obtained respectively by fitting the rough data [black circles in Fig.~\ref{fig.9}(c)] and the smooth data in Couette geometry [where a larger range of shear rates was accessible, see Fig.~\ref{fig.11}(a)] suggest that boundary conditions do have an impact on $\sigma_m$ whatever the applied shear rate.

In order to get a deeper insight on the stress overshoot with smooth boundary conditions, Fig.~\ref{fig.10}(a) focuses on three stress responses normalized by the maximum stress and plotted against the strain. It clearly appears that, while the stress response for $\dot\gamma=1$~s$^{-1}>\dot\gamma_s$ closely resembles that observed with rough boundary conditions see [Fig.~\ref{fig.3}(top)], stress responses for $\dot\gamma<\dot\gamma_s\simeq 0.2$~s$^{-1}$ show more complex features. For $\dot\gamma=0.1$~s$^{-1}\lesssim\dot\gamma_s$, the stress maximum is reached at a strain of about 0.4, i.e. well below 1. Moreover, after the maximum, the stress remains close to $\sigma_m$ and even increases at large strains. For $\dot\gamma=3.10^{-3}$~s$^{-1}$, the shape of $\sigma(t)$ is even more complex with a global maximum reached at $\gamma_m\simeq 0.25$ and several secondary maxima. We conclude that the characteristic shear rate $\dot\gamma_s$ separates two regimes where the stress responses are qualitatively different for low surface roughnesses: above $\dot\gamma_s$, a simple stress overshoot is observed as for rough boundary conditions whereas below $\dot\gamma_s$, more complex stress responses are recorded.

Since complex phenomena seem to be at play at low shear rates, we compare the stress responses $\sigma(t)$ under three different boundary conditions at the same shear rate $\dot\gamma= 5.10^{-4}$~s$^{-1}$ in Fig.~\ref{fig.10}(b). As already noted above, the failure strains are seen to decrease with decreasing roughness. Moreover, the stress response turns from smooth with rough walls to more erratic and fluctuating with smooth walls. This confirms the strong influence of boundary conditions on the failure scenario for $\dot\gamma<\dot\gamma_s$. Finally, the fluctuations observed with glass plates are reminiscent of stick-slip events \cite{Pignon:1996}. A full investigation of the local behaviour of the microgel in this very low shear regime will be the subject of a future study.

\subsubsection{Velocity profiles in a smooth Couette cell.}~As shown above, stress overshoots with smooth walls and at small shear rates ($\dot\gamma<\dot\gamma_s$) strongly differ from those observed with rough walls. Therefore, to elucidate the failure mechanism and the origin of this difference, we turn to USV measurements in a polished Plexiglas Couette cell. Global characteristics of the corresponding stress overshoots are gathered in Fig.~\ref{fig.11}(a) and (b). The same features as those already mentioned for the smooth plate-and-plate cell are observed, in particular the power-law regime of $\sigma_m$ with an exponent of 0.26 and the abrupt crossover at $\dot\gamma_s \simeq 0.2$~s$^{-1}$ between two decreasing bran\-ches in $t_m$. Additionally, the steady-state flow curve $\sigma$ vs $\dot \gamma$ of the microgel is drawn in Fig.~\ref{fig.11}(c). As reported in previous works \cite{Magnin:1990,Meeker:2004a,Meeker:2004b,Divoux:2010}, the flow curve deviates from the Herschel-Bulkley model at low shear rates. This behaviour is usually interpreted as the consequence of wall slip: in smooth cells and close to the yield stress of the material, rheological measurements are impaired by slippage at the wall, thus probing the rheology of lubrication layers rather than bulk properties. Here, it is quite remarkable that the characteristic shear rate $\dot\gamma_s$ inferred from transient overshoot data corresponds to the shear rate below which the flow curve shows a significant effect of wall slip.

\begin{figure*}[!t]\tt
\centering
\includegraphics[width=0.6\linewidth]{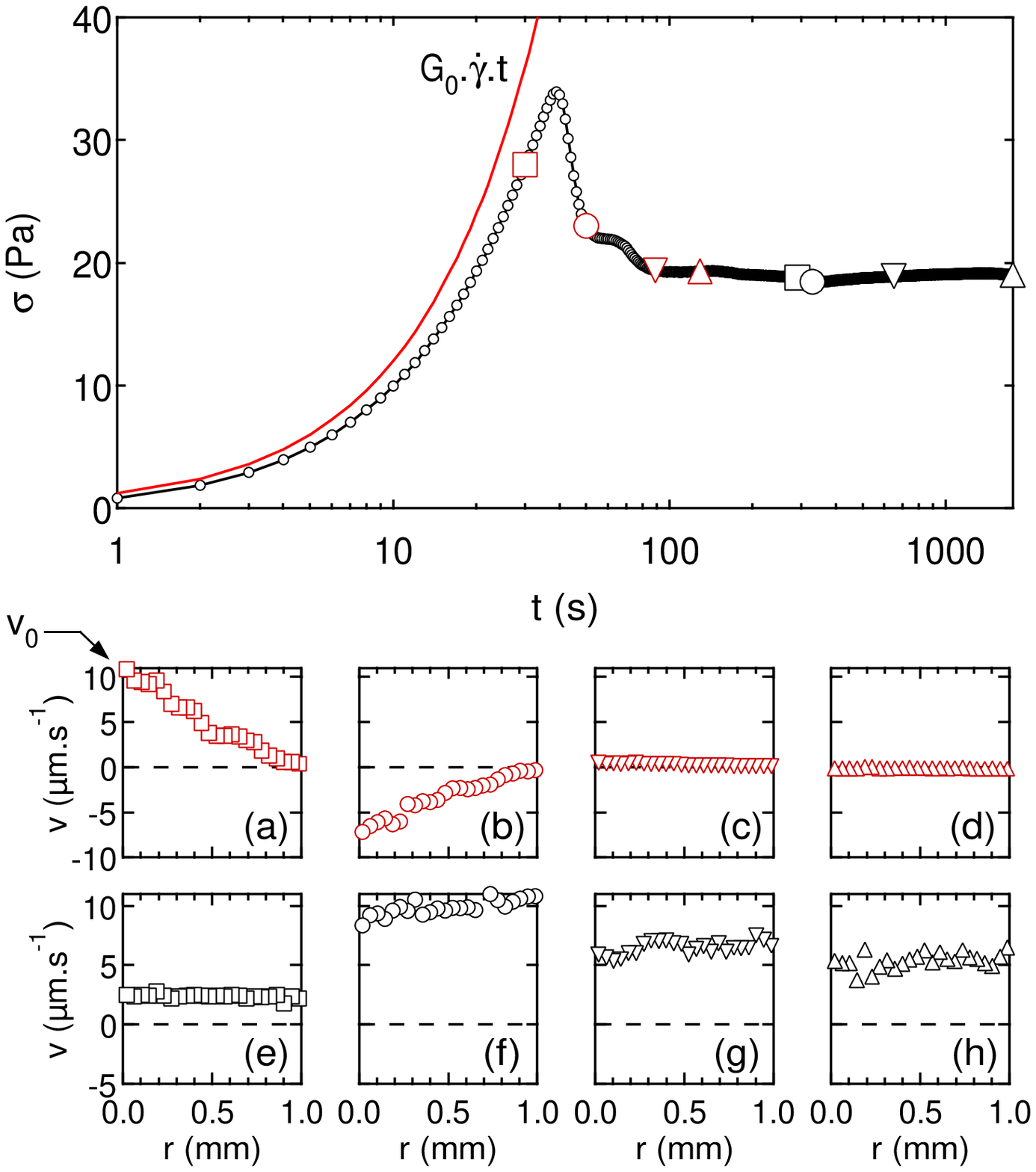}
\caption{Top: Shear stress $\sigma$ vs time $t$ for a shear rate $\dot \gamma = 0.01$~s$^{-1}$ applied at $t=0$. The red line shows $\sigma(t)=G_0\dot\gamma t$, where $G_0=120$~Pa is measured prior to shear start-up. Bottom: Velocity profiles $v(r)$ , where $r$ is the distance to the rotor, at different times [(letter), symbol, time (s)]: [(a), $\square$, 30]; [(b), $\circ$, 50]; [(c), $\triangledown$, 89]; [(d), $\vartriangle$, 129]; [(e), $\square$, 290]; [(f), $\circ$, 329]; [(g), $\vartriangle$, 649]; [(h), $\triangledown$, 1749]. The rotor velocity $v_0=10~\mu$m.s$^{-1}$ is indicated by an arrow. Same experimental conditions as in Fig.~\ref{fig.11}.}
\label{fig.12}
\end{figure*}

Figure~\ref{fig.12} shows the analysis of a typical stress overshoot for $\dot \gamma=0.01$~s$^{-1} < \dot\gamma_s$ using the same approach as in Fig.~\ref{fig.3}. As for rough boundary conditions, the initial growth of the stress corresponds to homogeneous strain [Fig.~\ref{fig.12}(a)] and the stress maximum also corresponds to failure at the rotor [Fig.~\ref{fig.12}(b)]. However, in the case of smooth boundary conditions, the failure of the microgel is followed by an elastic recoil that does not involve any subsequent temporal oscillations of the velocity field. Rather, the velocity immediately vanishes everywhere across the gap [Fig.~\ref{fig.12}(c)-(d)]. No motion is detected until $t\simeq 290$~s: at that time, slippage changes from the rotor side to the stator side within a few 10~s so that the velocity reaches the rotor velocity $v_0$ everywhere across the gap [Fig.~\ref{fig.12}(e)-(f)]. Finally, the velocity of this pluglike flow slowly decays to reach $v\simeq v_0/2$ at later times [Fig.~\ref{fig.12}(g)-(h)]. As for rough boundary conditions, we recall that, at least for $\dot \gamma > \dot\gamma_s$, such a pluglike flow later evolves toward a homogeneous flow through transient shear banding on much longer timescales \cite{Divoux:2010}. Therefore, although data on long enough timescales are not yet available for such a low shear rate as 0.01~s$^{-1}$, it is most likely that the pluglike flow of Fig.~\ref{fig.12}(h) does not represent the steady-state behaviour of the microgel. 

The velocity of the microgel close to the rotor is shown as a function of strain in Fig.~\ref{fig.13} (black line and symbols) together with velocity data from two other shear rates below $\dot\gamma_s$. The inset of Fig.~\ref{fig.13} clearly demonstrates that the microgel failure occurs at the same strain $\gamma_m\simeq 0.5$ whatever the shear rate and is followed by a very strong recoil without any velocity oscillation [compare with inset of Fig.~\ref{fig.4}(a)]. At large strains $\gamma\gtrsim 10$, all velocity signals are seen to converge toward $v\simeq v_0/2$, indicative of a pluglike flow with the same amount of wall slip at both walls. However, the transition from no-flow to pluglike flow at $v_0/2$ does not follow the same process for the three shear rates shown in Fig.~\ref{fig.13}: for $\dot \gamma=0.03$~s$^{-1}$, the velocity goes directly from 0 to $v_0/2$ at $\gamma\simeq 1.5$ whereas for the two other shear rates, wall slip at the rotor first totally disappears (so that $v=v_0$) before slowly increasing again (so that $v$ slowly decreases toward $v\simeq v_0/2$). Since we could not extract any clear correlation between the time evolution of these pluglike velocity profiles and the stress response $\sigma(t)$, we suggest that the behaviour of the microgel in smooth cells may also be heterogeneous in the vorticity direction as already observed in thixotropic laponite suspensions \cite{Gibaud:2008}. Once again, this regime will be the subject of a more detailed forthcoming work.

\section{Discussion and open questions}
\label{discuss}

\subsection{Aging and the stress overshoot}

Let us first come back to the evolution of the stress maximum $\sigma_m$ with the applied shear rate. Figure~\ref{fig.7} shows that for $\dot \gamma <\dot\gamma_w$ $\sigma_m$ remains constant or decreases with the shear rate while it increases as a weak power law for $\dot \gamma >\dot\gamma_w$. In fact, the decrease of $\sigma_m$ at low shear rate strongly depends on the value of the waiting time $t_w$. For a given small value of $t_w$, say $t_w\lesssim 60$~s, the elastic modulus $G'$ still significantly increases when shear is applied (see Fig.~\ref{fig.0}). Thus, a strong interplay between the
consolidation of the gel and the shear-induced fluidization is expected: the lower the applied shear rate, the larger the influence of the consolidation, leading to higher values of the stress maximum. This qualitatively explains the decrease of $\sigma_m$ with $\dot \gamma$. This decreasing trend disappears if shear is applied to an ``older'' microgel (i.e. for higher values of $t_w$): in this case [see Fig.~\ref{fig.6}(b-c)], the slow logarithmic growth of $G'$ indicates that the consolidation of the gel becomes negligible and should not play any role on the stress maximum which is thus independent of the applied shear rate. In other words, for large enough waiting times $t_w$, when $\dot \gamma <\dot\gamma_w$, the microgel behaves as a non-aging viscoelastic fluid which has ``forgotten'' its past preshear history.

For $\dot \gamma >\dot\gamma_w$, the growth of both $\sigma_m$ and $\gamma_m$ with $\dot \gamma$  indicates that the number of plastic events during the initial load is not independent of $\dot\gamma$ but rather decreases with $\dot\gamma$. As a consequence, the material gets effectively stiffer with $\dot\gamma$ and yielding is achieved at larger strains and stresses \cite{Derec:2003,Varnik:2004}. This trend is observed for both rough [Fig.~\ref{fig.6}(a)] and smooth boundary conditions [Fig.~\ref{fig.11}(a)] which confirms that this phenomenon is an intrinsic property of the fluid.
 
Interestingly, our results also show that for short waiting times $t_w$, a simple YSF may exhibit effects typical of aging, which is usually taken as the hallmark of thixotropic YSF. Therefore, the boundary between the two categories of YSF recalled in the introduction, i.e. simple and thixotropic YSF, is not as sharp as described in Refs.~\cite{Ragouilliaux:2007,Moller:2009b,Coussot:2010}. Although probably valid in steady state, such a distinction becomes less relevant when transient regimes are considered since the additional timescale $\dot\gamma^{-1}$ set by the imposed shear rate has to be taken into account. In practice, this raises the question of how to decide when a ``transient'' has died out and steady state is reached, which is particularly crucial for slow flows close to the yield stress.

\begin{figure}\tt
\centering
\includegraphics[width=0.9\columnwidth]{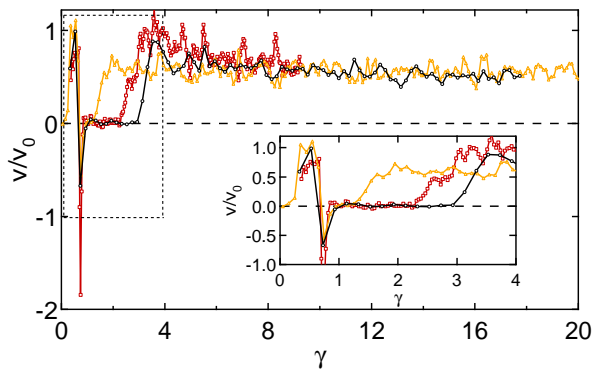}
\caption{Velocity $v(r=50~\mu$m$,t)$ normalized by the rotor velocity $v_0$ and plotted against the strain $\gamma=\dot\gamma t$ for various shear rates $\dot \gamma = 3.10^{-3}$ (red squares), $0.01$ (black circles) and 0.03~s$^{-1}$ (orange triangles) applied at $t=0$. $v$ is averaged over $\pm 50~\mu$m around the mean position $r=50~\mu$m from the rotor. Inset: horizontal zoom over the first four strain units which emphasizes the transition from no flow to pluglike flow after the microgel failure. Same experimental conditions as in Fig.~\ref{fig.11}.}
\label{fig.13}
\end{figure}

\subsection{Comparison with viscoelastic fluids}

Having emphasized the key role of both intrinsic timescales of the material and flow timescales in the dependence of the stress maximum with the imposed shear rate, one may wonder if the physical origin of the stress overshoot in YSF is the same as for viscoelastic fluids which do not present any yield stress such as wormlike micelle solutions, polymer melts, or entangled polymer solutions. In this last class of materials, a stress overshoot is usually observed only for shear rates larger than the inverse of the typical viscoelastic time of the microscopic constituents \cite{Berret:1997b,Yamamoto:2004}. However, here, stress overshoots in carbopol microgels are observed down to the smallest achievable shear rate of $10^{-4}$~s$^{-1}$. We can further point out two other important differences.

First, recent numerical work based on an elasto-plastic model built to describe foam rheology clearly showed that in the limit of low but finite yield strain, the stress overshoot disappears \cite{Raufaste:2010}. Such a result suggests that the existence of a stress overshoot in YSF is utterly related to the value of the yield stress. Second, a hallmark of stress overshoots in viscoelastic materials is the existence of a transient shear-banding phe\-no\-me\-non that occurs as soon as $\sigma_m$ is reached and persists when during the stress decrease. Such shear-banded flows concomittant with the stress overshoot have been observed in entangled polymer solutions \cite{Ravindranath:2008a,Boukany:2009b}, linear polymer melts \cite{Boukany:2009a}, and wormlike micelle solutions \cite{Lerouge:1998,Lerouge:2000}. They have also been recently predicted using the diffusive Rolie-Poly model \cite{Adams:2010}. In the case of carbopol microgels, the stress overshoot is followed by total wall slip and no shear banding is observed during the stress decrease at least for strains smaller than 10 (see Figs.~\ref{fig.3} and ~\ref{fig.13}). On much longer timescales, however, we indeed observe transient shear banding, as shown in Ref.~\cite{Divoux:2010} and in Fig.~\ref{fig.4}(b), leading to a homogeneous flow in steady state.

The only way to reconcile the two transient shear-banding phenomena would be to interpret the lubricating layers involved in slip flows as very thin shear bands smaller than the USV resolution and composed of fluidized carbopol\cite{Meeker:2004b}. However, if one assumes the width of the shear band to coincide with the USV resolution of 40~$\mu$m and based on the data of Fig.~\ref{fig.3}, one can estimate the local shear rate close to the rotor as $\sim 2.5\,10^3$~s$^{-1}$. With a shear stress of $\sim 40$~Pa, this corresponds to a local viscosity of 16~mPa.s within the shear band, which is more than four orders of magnitude smaller than the viscosity of fluidized carbopol ($\sim 400$~Pa.s here). Therefore, the layer close to the rotor cannot be interpreted as a very thin shear band. It is rather a true lubricating layer composed mostly of water, much thinner than 40~$\mu$m and possibly containing a few free carbopol particles. We conclude that in the case of our simple YSF, the stress overshoot is not directly related to shear banding as in viscoelastic fluids. The stress overshoot rather corresponds to failure at the wall. As seen in other soft matter systems \cite{Gibaud:2008,Lettinga:2009}, wall slip appears an alternative to shear banding for releasing stress in the bulk material. On longer timescales, we speculate that the highly-sheared lubrication layer progressively erodes the neighboring microgel through viscous friction. This leads to the slow growth of a fluidized region from the rotor, i.e. to the coexistence of two shear bands most probably presenting two different microstructures (two different levels of entanglement and/or volume fractions) of the soft spheres. Such a scenario could also explain why shear banding develops on timescales that can be orders of magnitude longer than that of the stress overshoot.

To conclude, the above major discrepancies between YSF and viscoelastic fluids suggest that the origin of the stress overshoot is different in the two cases. Some clues may eventually be found in the recent numerical calculations by Moorcroft {\it et al.} \cite{Moorcroft:2011} who compared predictions of a fluidity model to a modified version of the soft glassy rheology (SGR) model \cite{Sollich:1997,Sollich:1998}. While the fluidity model predicts shear-banded velocity profiles right after the stress overshoot and is probably not relevant to our experimental results, the modified SGR is able to capture a transient shear-banding phenomenon that develops on very long timescales for very old samples. Including the effects of wall roughness into such a model appears as the next step towards more quantitative comparisons with the experiments. 

\subsection{Exponents of the power-laws for $\sigma_m$ vs $\dot\gamma$}

Finally, the power-law dependence of the stress overshoot maximum with the shear rate raises three remaining open questions. (i) What are the reasons for the discrepancies between the experimental values of the exponent $\nu$ observed here for carbopol gels ($0.13<\nu<0.26$) on the one hand, and the theoretical exponent closer to $0.5$ observed for both a fluidity model \cite{Derec:2003} and Brownian dynamics simulations \cite{Whittle:1997} on the other hand? (ii) What is the origin of the difference between the power-law increase of $\sigma_m$ reported in this article and the logarithmic behaviour observed in binary Lennard-Jones mixtures under shear \cite{Varnik:2004,Rottler:2005}? In the latter case, it would be interesting to test the relevance of weak power laws to describe the evolution of the stress maximum with the strain rate. (iii) What is the control parameter for the value of the exponent $\nu$ measured in the experiments? Is this value characteristic of the underlying microstructure, with attractive systems leading to larger exponents than repulsive ones \cite{Derec:2003,Koumakis:2011}?

\section{Summary and outlook}

The stress overshoot is a widespread phenomenon observed in various systems. In this article, we have reported an extensive data set on the stress overshoot in a simple yield stress fluid during start-up experiments under imposed shear rate. The local scenario of the stress overshoot phenomenon is as follows: (i) at small strains ($\gamma \lesssim 1$), the microgel undergoes homogeneous deformation hinting to elasto-plastic behaviour; (ii) the maximum stress $\sigma_m$ corresponds to the failure of the microgel and to the nucleation of a thin lubrication layer at the moving wall; (iii) the microgel then experiences a strong elastic recoil and enters a regime of total wall slip while the stress decreases. This scenario is true for both rough boundary conditions for all explored shear rates ($10^{-4}<\dot\gamma<10$~s$^{-1}$) and for smooth BC at high enough shear rates ($\dot\gamma > \dot\gamma_s$). In this latter case, the local scenario is modified at low applied shear rates ($\dot\gamma < \dot\gamma_s$): failure occurs at smaller strains and slippage may be observed at both walls together with fluctuating stress responses reminiscent of stick-slip behaviour.

We have shown that the stress maximum $\sigma_m$ reached during the overshoot is roughly constant at low applied shear rates $\dot\gamma<\dot\gamma_w$ and increases as a weak power law for $\dot\gamma>\dot\gamma_w$, where $\dot\gamma_w$ scales as the inverse of the waiting time $t_w$ between the preshear and the start of the experiment. Such a dependence is very robust over a large range of carbopol concentrations and boundary conditions. The exponent $\nu$ only slightly depends on the batch preparation and on the carbopol concentration. However, it seems to depend significantly on the boundary conditions and its value ranges from 0.13 to 0.26 over the range of parameters explored in the present study. 

Future work will deal with both surface and bulk behaviours of carbopol microgels in the limit of very low applied shear rates under smooth boundary conditions. Indeed, the erratic nature of the stress response in these conditions [see Fig.~\ref{fig.10}~(b)] still remains to be fully elucidated. In particular, it would be of great interest to locate plastic rearrangements in the bulk and to correlate them with stress fluctuations, so as to disclose the analogies and the differences between such a stick-slip-like behaviour and the one observed in recent numerical simulations of Lennard-Jones amorphous solids \cite{Tanguy:2006,Tsamados:2010}. The fact that such complex temporal behaviours arise for applied shear rates lower than that where the steady-state flow curve presents a kink attributed to wall slip [see Fig.~\ref{fig.11}(c)] also certainly deserves further investigation. Finally, the differences between the stress overshoot reported here and that observed in viscoelastic polymerlike solutions remain to be fully understood. In particular, we have shown that in our carbopol microgels shear banding sets in over timescales much longer than that of the stress overshoot, while in polymer solutions and melts, the duration of the transient shear-banding regime is of the same order as the stress overshoot duration. In both cases, a complete microscopic interpretation is still needed.

\begin{acknowledgments}
We thank Y. Forterre for providing us with the carbopol, D.~Tamarii for substantial help with the experiments and with the software, and H.~Feret for technical help. M. Cloitre is deeply thanked for enlightening discussions as well as P. Chaudhuri, A. Lemaitre, G. Ovarlez, A .Tanguy and D. Weaire.
\end{acknowledgments}

\bibliography{bibseb} 

\providecommand*{\mcitethebibliography}{\thebibliography}
\csname @ifundefined\endcsname{endmcitethebibliography}
{\let\endmcitethebibliography\endthebibliography}{}
\begin{mcitethebibliography}{96}
\providecommand*{\natexlab}[1]{#1}
\providecommand*{\mciteSetBstSublistMode}[1]{}
\providecommand*{\mciteSetBstMaxWidthForm}[2]{}
\providecommand*{\mciteBstWouldAddEndPuncttrue}
  {\def\EndOfBibitem{\unskip.}}
\providecommand*{\mciteBstWouldAddEndPunctfalse}
  {\let\EndOfBibitem\relax}
\providecommand*{\mciteSetBstMidEndSepPunct}[3]{}
\providecommand*{\mciteSetBstSublistLabelBeginEnd}[3]{}
\providecommand*{\EndOfBibitem}{}
\mciteSetBstSublistMode{f}
\mciteSetBstMaxWidthForm{subitem}
{(\emph{\alph{mcitesubitemcount}})}
\mciteSetBstSublistLabelBeginEnd{\mcitemaxwidthsubitemform\space}
{\relax}{\relax}

\bibitem[Arruda \emph{et~al.}(1995)Arruda, Boyce, and
  Jayachandran]{Arruda:1995}
E.~M. Arruda, M.~C. Boyce and R.~Jayachandran, \emph{Mech. Mater.}, 1995,
  \textbf{19}, 193--212\relax
\mciteBstWouldAddEndPuncttrue
\mciteSetBstMidEndSepPunct{\mcitedefaultmidpunct}
{\mcitedefaultendpunct}{\mcitedefaultseppunct}\relax
\EndOfBibitem
\bibitem[van Melick \emph{et~al.}(2003)van Melick, Govaert, and
  Meijer]{Melick:2003}
H.~G.~H. van Melick, L.~E. Govaert and H.~E.~H. Meijer, \emph{Polymer}, 2003,
  \textbf{44}, 3579--3591\relax
\mciteBstWouldAddEndPuncttrue
\mciteSetBstMidEndSepPunct{\mcitedefaultmidpunct}
{\mcitedefaultendpunct}{\mcitedefaultseppunct}\relax
\EndOfBibitem
\bibitem[van Melick \emph{et~al.}(2003)van Melick, Govaert, and
  Meijer]{Melick:2003b}
H.~G.~H. van Melick, L.~E. Govaert and H.~E.~H. Meijer, \emph{Polymer}, 2003,
  \textbf{44}, 457--465\relax
\mciteBstWouldAddEndPuncttrue
\mciteSetBstMidEndSepPunct{\mcitedefaultmidpunct}
{\mcitedefaultendpunct}{\mcitedefaultseppunct}\relax
\EndOfBibitem
\bibitem[van Aken \emph{et~al.}(2000)van Aken, de~Hey, and Sietsma]{Aken:2000}
B.~van Aken, P.~de~Hey and J.~Sietsma, \emph{Mater. Sci. Eng. A}, 2000,
  \textbf{278}, 247--254\relax
\mciteBstWouldAddEndPuncttrue
\mciteSetBstMidEndSepPunct{\mcitedefaultmidpunct}
{\mcitedefaultendpunct}{\mcitedefaultseppunct}\relax
\EndOfBibitem
\bibitem[Johnson \emph{et~al.}(2002)Johnson, Lu, and Demetrios]{Johnson:2002}
W.~L. Johnson, J.~Lu and M.~D. Demetrios, \emph{Intermetallics}, 2002,
  \textbf{10}, 1039--1046\relax
\mciteBstWouldAddEndPuncttrue
\mciteSetBstMidEndSepPunct{\mcitedefaultmidpunct}
{\mcitedefaultendpunct}{\mcitedefaultseppunct}\relax
\EndOfBibitem
\bibitem[Partal \emph{et~al.}(1999)Partal, Guerrero, Berjano, and
  Gallegos]{Partal:1999}
P.~Partal, A.~Guerrero, M.~Berjano and C.~Gallegos, \emph{J. Food Eng.}, 1999,
  \textbf{41}, 33--41\relax
\mciteBstWouldAddEndPuncttrue
\mciteSetBstMidEndSepPunct{\mcitedefaultmidpunct}
{\mcitedefaultendpunct}{\mcitedefaultseppunct}\relax
\EndOfBibitem
\bibitem[B{\'e}cu \emph{et~al.}(2005)B{\'e}cu, Grondin, Manneville, and
  Colin]{Becu:2005}
L.~B{\'e}cu, P.~Grondin, S.~Manneville and A.~Colin, \emph{Colloids Surfaces
  A}, 2005, \textbf{263}, 146--152\relax
\mciteBstWouldAddEndPuncttrue
\mciteSetBstMidEndSepPunct{\mcitedefaultmidpunct}
{\mcitedefaultendpunct}{\mcitedefaultseppunct}\relax
\EndOfBibitem
\bibitem[B{\'e}cu \emph{et~al.}(2006)B{\'e}cu, Manneville, and
  Colin]{Becu:2006}
L.~B{\'e}cu, S.~Manneville and A.~Colin, \emph{Phys. Rev. Lett.}, 2006,
  \textbf{96}, 138302\relax
\mciteBstWouldAddEndPuncttrue
\mciteSetBstMidEndSepPunct{\mcitedefaultmidpunct}
{\mcitedefaultendpunct}{\mcitedefaultseppunct}\relax
\EndOfBibitem
\bibitem[Ovarlez \emph{et~al.}(2008)Ovarlez, Rodts, Ragouilliaux, Coussot,
  Goyon, and Colin]{Ovarlez:2008}
G.~Ovarlez, S.~Rodts, A.~Ragouilliaux, P.~Coussot, J.~Goyon and A.~Colin,
  \emph{Phys. Rev. E}, 2008, \textbf{78}, 036307\relax
\mciteBstWouldAddEndPuncttrue
\mciteSetBstMidEndSepPunct{\mcitedefaultmidpunct}
{\mcitedefaultendpunct}{\mcitedefaultseppunct}\relax
\EndOfBibitem
\bibitem[Khan \emph{et~al.}(1988)Khan, Schnepper, and Armstrong]{Khan:1988}
S.~A. Khan, C.~A. Schnepper and R.~C. Armstrong, \emph{J. Rheol.}, 1988,
  \textbf{32}, 69--92\relax
\mciteBstWouldAddEndPuncttrue
\mciteSetBstMidEndSepPunct{\mcitedefaultmidpunct}
{\mcitedefaultendpunct}{\mcitedefaultseppunct}\relax
\EndOfBibitem
\bibitem[Raufaste \emph{et~al.}(2010)Raufaste, Cox, Marmottant, and
  Graner]{Raufaste:2010}
C.~Raufaste, S.~J. Cox, P.~Marmottant and F.~Graner, \emph{Phys. Rev. E}, 2010,
  \textbf{81}, 031404\relax
\mciteBstWouldAddEndPuncttrue
\mciteSetBstMidEndSepPunct{\mcitedefaultmidpunct}
{\mcitedefaultendpunct}{\mcitedefaultseppunct}\relax
\EndOfBibitem
\bibitem[Islam \emph{et~al.}(2004)Islam, Rodriguez-Hornedo, Ciotti, and
  Ackermann]{Islam:2004}
M.~T. Islam, N.~Rodriguez-Hornedo, S.~Ciotti and C.~Ackermann, \emph{Pharm.
  Res.}, 2004, \textbf{21}, 1192--1199\relax
\mciteBstWouldAddEndPuncttrue
\mciteSetBstMidEndSepPunct{\mcitedefaultmidpunct}
{\mcitedefaultendpunct}{\mcitedefaultseppunct}\relax
\EndOfBibitem
\bibitem[Coussot \emph{et~al.}(2009)Coussot, Tocquer, Lanos, and
  Ovarlez]{Coussot:2009}
P.~Coussot, L.~Tocquer, C.~Lanos and G.~Ovarlez, \emph{J. Non-Newtonian Fluid
  Mech.}, 2009, \textbf{158}, 85--90\relax
\mciteBstWouldAddEndPuncttrue
\mciteSetBstMidEndSepPunct{\mcitedefaultmidpunct}
{\mcitedefaultendpunct}{\mcitedefaultseppunct}\relax
\EndOfBibitem
\bibitem[Wroth(1958)]{Wroth:1958}
C.~Wroth, \emph{Engineering}, 1958, \textbf{186}, 409--413\relax
\mciteBstWouldAddEndPuncttrue
\mciteSetBstMidEndSepPunct{\mcitedefaultmidpunct}
{\mcitedefaultendpunct}{\mcitedefaultseppunct}\relax
\EndOfBibitem
\bibitem[Geminard \emph{et~al.}(1999)Geminard, Losert, and
  Gollub]{Geminard:1999}
J.-C. Geminard, W.~Losert and J.-P. Gollub, \emph{Phys. Rev. E}, 1999,
  \textbf{59}, 5881--5890\relax
\mciteBstWouldAddEndPuncttrue
\mciteSetBstMidEndSepPunct{\mcitedefaultmidpunct}
{\mcitedefaultendpunct}{\mcitedefaultseppunct}\relax
\EndOfBibitem
\bibitem[Letwimolnun \emph{et~al.}(2007)Letwimolnun, Vergnes, Ausias, and
  Carreau]{Letwimolnun:2007}
W.~Letwimolnun, B.~Vergnes, G.~Ausias and P.~J. Carreau, \emph{J. Non-Newtonian
  Fluid Mech.}, 2007, \textbf{141}, 167--179\relax
\mciteBstWouldAddEndPuncttrue
\mciteSetBstMidEndSepPunct{\mcitedefaultmidpunct}
{\mcitedefaultendpunct}{\mcitedefaultseppunct}\relax
\EndOfBibitem
\bibitem[Nguyen and Boger(1983)]{Nguyen:1983}
Q.~D. Nguyen and D.~V. Boger, \emph{J. Rheol.}, 1983, \textbf{27},
  321--349\relax
\mciteBstWouldAddEndPuncttrue
\mciteSetBstMidEndSepPunct{\mcitedefaultmidpunct}
{\mcitedefaultendpunct}{\mcitedefaultseppunct}\relax
\EndOfBibitem
\bibitem[Persello \emph{et~al.}(1994)Persello, Magnin, Chang, Piau, and
  Cabane]{Persello:1994}
J.~Persello, A.~Magnin, J.~Chang, J.~Piau and B.~Cabane, \emph{J. Rheol.},
  1994, \textbf{38}, 1845--1870\relax
\mciteBstWouldAddEndPuncttrue
\mciteSetBstMidEndSepPunct{\mcitedefaultmidpunct}
{\mcitedefaultendpunct}{\mcitedefaultseppunct}\relax
\EndOfBibitem
\bibitem[Derec \emph{et~al.}(2003)Derec, Ducouret, Ajdari, and
  Lequeux]{Derec:2003}
C.~Derec, G.~Ducouret, A.~Ajdari and F.~Lequeux, \emph{Phys. Rev. E}, 2003,
  \textbf{67}, 061403\relax
\mciteBstWouldAddEndPuncttrue
\mciteSetBstMidEndSepPunct{\mcitedefaultmidpunct}
{\mcitedefaultendpunct}{\mcitedefaultseppunct}\relax
\EndOfBibitem
\bibitem[Mahaut \emph{et~al.}(2008)Mahaut, Chateau, Coussot, and
  Ovarlez]{Mahaut:2008}
F.~Mahaut, X.~Chateau, P.~Coussot and G.~Ovarlez, \emph{J. Rheol.}, 2008,
  \textbf{52}, 287--313\relax
\mciteBstWouldAddEndPuncttrue
\mciteSetBstMidEndSepPunct{\mcitedefaultmidpunct}
{\mcitedefaultendpunct}{\mcitedefaultseppunct}\relax
\EndOfBibitem
\bibitem[Koumakis and Petekidis(2011)]{Koumakis:2011}
N.~Koumakis and G.~Petekidis, \emph{Soft Matter}, 2011, \textbf{7},
  2456--2470\relax
\mciteBstWouldAddEndPuncttrue
\mciteSetBstMidEndSepPunct{\mcitedefaultmidpunct}
{\mcitedefaultendpunct}{\mcitedefaultseppunct}\relax
\EndOfBibitem
\bibitem[Rottler and Robbins(2003)]{Rottler:2003}
J.~Rottler and M.~O. Robbins, \emph{Phys. Rev. E}, 2003, \textbf{68},
  011507\relax
\mciteBstWouldAddEndPuncttrue
\mciteSetBstMidEndSepPunct{\mcitedefaultmidpunct}
{\mcitedefaultendpunct}{\mcitedefaultseppunct}\relax
\EndOfBibitem
\bibitem[Varnik \emph{et~al.}(2004)Varnik, Bocquet, and Barrat]{Varnik:2004}
F.~Varnik, L.~Bocquet and J.-L. Barrat, \emph{J. Chem. Phys.}, 2004,
  \textbf{120}, 2788--2801\relax
\mciteBstWouldAddEndPuncttrue
\mciteSetBstMidEndSepPunct{\mcitedefaultmidpunct}
{\mcitedefaultendpunct}{\mcitedefaultseppunct}\relax
\EndOfBibitem
\bibitem[Rottler and Robbins(2005)]{Rottler:2005}
J.~Rottler and M.~O. Robbins, \emph{Phys. Rev. Lett.}, 2005, \textbf{95},
  225504\relax
\mciteBstWouldAddEndPuncttrue
\mciteSetBstMidEndSepPunct{\mcitedefaultmidpunct}
{\mcitedefaultendpunct}{\mcitedefaultseppunct}\relax
\EndOfBibitem
\bibitem[Langer and Pechenik(2003)]{Langer:2003}
J.~S. Langer and L.~Pechenik, \emph{Phys. Rev. E}, 2003, \textbf{68},
  061507\relax
\mciteBstWouldAddEndPuncttrue
\mciteSetBstMidEndSepPunct{\mcitedefaultmidpunct}
{\mcitedefaultendpunct}{\mcitedefaultseppunct}\relax
\EndOfBibitem
\bibitem[Jagla(2007)]{Jagla:2007}
E.~A. Jagla, \emph{Phys. Rev. E}, 2007, \textbf{76}, 046119\relax
\mciteBstWouldAddEndPuncttrue
\mciteSetBstMidEndSepPunct{\mcitedefaultmidpunct}
{\mcitedefaultendpunct}{\mcitedefaultseppunct}\relax
\EndOfBibitem
\bibitem[Jagla(2010)]{Jagla:2010}
E.~A. Jagla, \emph{J. Stat. Mech.}, 2010,  P12025\relax
\mciteBstWouldAddEndPuncttrue
\mciteSetBstMidEndSepPunct{\mcitedefaultmidpunct}
{\mcitedefaultendpunct}{\mcitedefaultseppunct}\relax
\EndOfBibitem
\bibitem[Falk and Langer(2010)]{Falk:2010pp}
M.~Falk and J.~Langer, E-print cond-mat/1004.4684\relax
\mciteBstWouldAddEndPuncttrue
\mciteSetBstMidEndSepPunct{\mcitedefaultmidpunct}
{\mcitedefaultendpunct}{\mcitedefaultseppunct}\relax
\EndOfBibitem
\bibitem[Sollich(1998)]{Sollich:1998}
P.~Sollich, \emph{Phys. Rev. E}, 1998, \textbf{58}, 738--759\relax
\mciteBstWouldAddEndPuncttrue
\mciteSetBstMidEndSepPunct{\mcitedefaultmidpunct}
{\mcitedefaultendpunct}{\mcitedefaultseppunct}\relax
\EndOfBibitem
\bibitem[Fielding \emph{et~al.}(2000)Fielding, Sollich, and
  Cates]{Fielding:2000}
S.~M. Fielding, P.~Sollich and M.~E. Cates, \emph{J. Rheol.}, 2000,
  \textbf{44}, 323--369\relax
\mciteBstWouldAddEndPuncttrue
\mciteSetBstMidEndSepPunct{\mcitedefaultmidpunct}
{\mcitedefaultendpunct}{\mcitedefaultseppunct}\relax
\EndOfBibitem
\bibitem[Moorcroft \emph{et~al.}(2011)Moorcroft, Cates, and
  Fielding]{Moorcroft:2011}
R.~L. Moorcroft, M.~E. Cates and S.~M. Fielding, \emph{Phys. Rev. Lett.}, 2011,
  \textbf{106}, 055502\relax
\mciteBstWouldAddEndPuncttrue
\mciteSetBstMidEndSepPunct{\mcitedefaultmidpunct}
{\mcitedefaultendpunct}{\mcitedefaultseppunct}\relax
\EndOfBibitem
\bibitem[Whittle and Dickinson(1997)]{Whittle:1997}
M.~Whittle and E.~Dickinson, \emph{J. Chem. Phys.}, 1997, \textbf{107},
  10191--10200\relax
\mciteBstWouldAddEndPuncttrue
\mciteSetBstMidEndSepPunct{\mcitedefaultmidpunct}
{\mcitedefaultendpunct}{\mcitedefaultseppunct}\relax
\EndOfBibitem
\bibitem[Albano \emph{et~al.}(2004)Albano, Lacevic, Falk, and
  Glotzer]{Albano:2004}
F.~Albano, N.~Lacevic, M.~L. Falk and S.~C. Glotzer, \emph{Mater. Sci. Eng. A},
  2004, \textbf{375--377}, 671--674\relax
\mciteBstWouldAddEndPuncttrue
\mciteSetBstMidEndSepPunct{\mcitedefaultmidpunct}
{\mcitedefaultendpunct}{\mcitedefaultseppunct}\relax
\EndOfBibitem
\bibitem[Xu and O'Hern(2006)]{Xu:2006}
N.~Xu and C.~S. O'Hern, \emph{Phys. Rev. E}, 2006, \textbf{73}, 061303\relax
\mciteBstWouldAddEndPuncttrue
\mciteSetBstMidEndSepPunct{\mcitedefaultmidpunct}
{\mcitedefaultendpunct}{\mcitedefaultseppunct}\relax
\EndOfBibitem
\bibitem[Zausch \emph{et~al.}(2008)Zausch, Horbach, Laurati, Egelhaaf, Brader,
  Voigtmann, and Fuchs]{Zausch:2008}
J.~Zausch, J.~Horbach, M.~Laurati, S.~U. Egelhaaf, J.~M. Brader, T.~Voigtmann
  and M.~Fuchs, \emph{J. Phys.: Condens. Matter}, 2008, \textbf{20},
  404210\relax
\mciteBstWouldAddEndPuncttrue
\mciteSetBstMidEndSepPunct{\mcitedefaultmidpunct}
{\mcitedefaultendpunct}{\mcitedefaultseppunct}\relax
\EndOfBibitem
\bibitem[Berret(1997)]{Berret:1997b}
J.-F. Berret, \emph{Langmuir}, 1997, \textbf{13}, 2227--2234\relax
\mciteBstWouldAddEndPuncttrue
\mciteSetBstMidEndSepPunct{\mcitedefaultmidpunct}
{\mcitedefaultendpunct}{\mcitedefaultseppunct}\relax
\EndOfBibitem
\bibitem[Lerouge \emph{et~al.}(1998)Lerouge, Decruppe, and
  Humbert]{Lerouge:1998}
S.~Lerouge, J.-P. Decruppe and C.~Humbert, \emph{Phys. Rev. Lett.}, 1998,
  \textbf{81}, 5457--5460\relax
\mciteBstWouldAddEndPuncttrue
\mciteSetBstMidEndSepPunct{\mcitedefaultmidpunct}
{\mcitedefaultendpunct}{\mcitedefaultseppunct}\relax
\EndOfBibitem
\bibitem[Soltero \emph{et~al.}(1999)Soltero, Bautista, Puig, and
  Manero]{Soltero:1999}
J.~F.~A. Soltero, F.~Bautista, J.~E. Puig and O.~Manero, \emph{Langmuir}, 1999,
  \textbf{15}, 1604--1612\relax
\mciteBstWouldAddEndPuncttrue
\mciteSetBstMidEndSepPunct{\mcitedefaultmidpunct}
{\mcitedefaultendpunct}{\mcitedefaultseppunct}\relax
\EndOfBibitem
\bibitem[Lerouge \emph{et~al.}(2000)Lerouge, Decruppe, and
  Berret]{Lerouge:2000}
S.~Lerouge, J.-P. Decruppe and J.-F. Berret, \emph{Langmuir}, 2000,
  \textbf{16}, 6464--6474\relax
\mciteBstWouldAddEndPuncttrue
\mciteSetBstMidEndSepPunct{\mcitedefaultmidpunct}
{\mcitedefaultendpunct}{\mcitedefaultseppunct}\relax
\EndOfBibitem
\bibitem[Decruppe \emph{et~al.}(2001)Decruppe, Lerouge, and
  Berret]{Decruppe:2001}
J.-P. Decruppe, S.~Lerouge and J.-F. Berret, \emph{Phys. Rev. E}, 2001,
  \textbf{63}, 022501\relax
\mciteBstWouldAddEndPuncttrue
\mciteSetBstMidEndSepPunct{\mcitedefaultmidpunct}
{\mcitedefaultendpunct}{\mcitedefaultseppunct}\relax
\EndOfBibitem
\bibitem[Lerouge \emph{et~al.}(2004)Lerouge, Decruppe, and
  Olmsted]{Lerouge:2004}
S.~Lerouge, J.-P. Decruppe and P.~Olmsted, \emph{Langmuir}, 2004, \textbf{20},
  11355--11365\relax
\mciteBstWouldAddEndPuncttrue
\mciteSetBstMidEndSepPunct{\mcitedefaultmidpunct}
{\mcitedefaultendpunct}{\mcitedefaultseppunct}\relax
\EndOfBibitem
\bibitem[Ganapathy and Sood(2008)]{Ganapathy:2008}
R.~Ganapathy and A.~K. Sood, \emph{J. Non-Newtonian Fluid Mech.}, 2008,
  \textbf{149}, 78--86\relax
\mciteBstWouldAddEndPuncttrue
\mciteSetBstMidEndSepPunct{\mcitedefaultmidpunct}
{\mcitedefaultendpunct}{\mcitedefaultseppunct}\relax
\EndOfBibitem
\bibitem[Yamamoto and Onuki(2004)]{Yamamoto:2004}
R.~Yamamoto and A.~Onuki, \emph{Phys. Rev. E}, 2004, \textbf{70}, 041801\relax
\mciteBstWouldAddEndPuncttrue
\mciteSetBstMidEndSepPunct{\mcitedefaultmidpunct}
{\mcitedefaultendpunct}{\mcitedefaultseppunct}\relax
\EndOfBibitem
\bibitem[Boukany and Wang(2009)]{Boukany:2009a}
P.~E. Boukany and S.-Q. Wang, \emph{J. Rheol.}, 2009, \textbf{53},
  617--629\relax
\mciteBstWouldAddEndPuncttrue
\mciteSetBstMidEndSepPunct{\mcitedefaultmidpunct}
{\mcitedefaultendpunct}{\mcitedefaultseppunct}\relax
\EndOfBibitem
\bibitem[Tapadia and Wang(2004)]{Tapadia:2004}
P.~Tapadia and S.-Q. Wang, \emph{Macromolecules}, 2004, \textbf{37},
  9083--9095\relax
\mciteBstWouldAddEndPuncttrue
\mciteSetBstMidEndSepPunct{\mcitedefaultmidpunct}
{\mcitedefaultendpunct}{\mcitedefaultseppunct}\relax
\EndOfBibitem
\bibitem[Ravindranath and Wang(2008)]{Ravindranath:2008b}
S.~Ravindranath and S.-Q. Wang, \emph{J. Rheol.}, 2008, \textbf{52},
  681--695\relax
\mciteBstWouldAddEndPuncttrue
\mciteSetBstMidEndSepPunct{\mcitedefaultmidpunct}
{\mcitedefaultendpunct}{\mcitedefaultseppunct}\relax
\EndOfBibitem
\bibitem[Boukany and Wang(2009)]{Boukany:2009b}
P.~E. Boukany and S.-Q. Wang, \emph{Soft Matter}, 2009, \textbf{5},
  780--789\relax
\mciteBstWouldAddEndPuncttrue
\mciteSetBstMidEndSepPunct{\mcitedefaultmidpunct}
{\mcitedefaultendpunct}{\mcitedefaultseppunct}\relax
\EndOfBibitem
\bibitem[Boukany \emph{et~al.}(2010)Boukany, Hemminger, Wang, and
  Lee]{Boukany:2010}
P.~E. Boukany, O.~Hemminger, S.-Q. Wang and L.~J. Lee, \emph{Phys. Rev. Lett.},
  2010, \textbf{105}, 027802\relax
\mciteBstWouldAddEndPuncttrue
\mciteSetBstMidEndSepPunct{\mcitedefaultmidpunct}
{\mcitedefaultendpunct}{\mcitedefaultseppunct}\relax
\EndOfBibitem
\bibitem[Tapadia and Wang(2006)]{Tapadia:2006a}
P.~Tapadia and S.-Q. Wang, \emph{Phys. Rev. Lett.}, 2006, \textbf{96},
  016001\relax
\mciteBstWouldAddEndPuncttrue
\mciteSetBstMidEndSepPunct{\mcitedefaultmidpunct}
{\mcitedefaultendpunct}{\mcitedefaultseppunct}\relax
\EndOfBibitem
\bibitem[Wang and Wang(2009)]{Wang:2009}
Y.~Wang and S.-Q. Wang, \emph{J. Rheol.}, 2009, \textbf{53}, 1389--1401\relax
\mciteBstWouldAddEndPuncttrue
\mciteSetBstMidEndSepPunct{\mcitedefaultmidpunct}
{\mcitedefaultendpunct}{\mcitedefaultseppunct}\relax
\EndOfBibitem
\bibitem[Ravindranath \emph{et~al.}(2008)Ravindranath, Wang, Olechnowicz, and
  Quirck]{Ravindranath:2008a}
S.~Ravindranath, S.-Q. Wang, M.~Olechnowicz and R.~P. Quirck,
  \emph{Macromolecules}, 2008, \textbf{41}, 2663--2670\relax
\mciteBstWouldAddEndPuncttrue
\mciteSetBstMidEndSepPunct{\mcitedefaultmidpunct}
{\mcitedefaultendpunct}{\mcitedefaultseppunct}\relax
\EndOfBibitem
\bibitem[Adams and Olmsted(2009)]{Adams:2009}
J.~M. Adams and P.~D. Olmsted, \emph{Phys. Rev. Lett.}, 2009, \textbf{102},
  067801\relax
\mciteBstWouldAddEndPuncttrue
\mciteSetBstMidEndSepPunct{\mcitedefaultmidpunct}
{\mcitedefaultendpunct}{\mcitedefaultseppunct}\relax
\EndOfBibitem
\bibitem[Adams \emph{et~al.}(2010)Adams, Fielding, and Olmsted]{Adams:2010}
J.~M. Adams, S.~M. Fielding and P.~D. Olmsted, E-print cond-mat/1011.4355\relax
\mciteBstWouldAddEndPuncttrue
\mciteSetBstMidEndSepPunct{\mcitedefaultmidpunct}
{\mcitedefaultendpunct}{\mcitedefaultseppunct}\relax
\EndOfBibitem
\bibitem[Nguyen and Boger(1992)]{Nguyen:1992}
Q.~D. Nguyen and D.~V. Boger, \emph{Annu. Rev. Fluid Mech.}, 1992, \textbf{24},
  47--88\relax
\mciteBstWouldAddEndPuncttrue
\mciteSetBstMidEndSepPunct{\mcitedefaultmidpunct}
{\mcitedefaultendpunct}{\mcitedefaultseppunct}\relax
\EndOfBibitem
\bibitem[Barnes(1999)]{Barnes:1999}
H.~A. Barnes, \emph{J. Non-Newtonian Fluid Mech.}, 1999, \textbf{81},
  133--178\relax
\mciteBstWouldAddEndPuncttrue
\mciteSetBstMidEndSepPunct{\mcitedefaultmidpunct}
{\mcitedefaultendpunct}{\mcitedefaultseppunct}\relax
\EndOfBibitem
\bibitem[M{\o}ller \emph{et~al.}(2009)M{\o}ller, Fall, and Bonn]{Moller:2009a}
P.~C.~F. M{\o}ller, A.~Fall and D.~Bonn, \emph{Europhys. Lett.}, 2009,
  \textbf{87}, 38004\relax
\mciteBstWouldAddEndPuncttrue
\mciteSetBstMidEndSepPunct{\mcitedefaultmidpunct}
{\mcitedefaultendpunct}{\mcitedefaultseppunct}\relax
\EndOfBibitem
\bibitem[Ragouilliaux \emph{et~al.}(2007)Ragouilliaux, Ovarlez,
  Shahidzadeh-Bonn, Herzhaft, Palermo, and Coussot]{Ragouilliaux:2007}
A.~Ragouilliaux, G.~Ovarlez, N.~Shahidzadeh-Bonn, B.~Herzhaft, T.~Palermo and
  P.~Coussot, \emph{Phys. Rev. E}, 2007, \textbf{76}, 051408\relax
\mciteBstWouldAddEndPuncttrue
\mciteSetBstMidEndSepPunct{\mcitedefaultmidpunct}
{\mcitedefaultendpunct}{\mcitedefaultseppunct}\relax
\EndOfBibitem
\bibitem[M{\o}ller \emph{et~al.}(2008)M{\o}ller, Rodts, Michels, and
  Bonn]{Moller:2008}
P.~C.~F. M{\o}ller, S.~Rodts, M.~A.~J. Michels and D.~Bonn, \emph{Phys. Rev.
  E}, 2008, \textbf{77}, 041507\relax
\mciteBstWouldAddEndPuncttrue
\mciteSetBstMidEndSepPunct{\mcitedefaultmidpunct}
{\mcitedefaultendpunct}{\mcitedefaultseppunct}\relax
\EndOfBibitem
\bibitem[Magnin and Piau(1990)]{Magnin:1990}
A.~Magnin and J.~Piau, \emph{J. Non-Newtonian Fluid Mech.}, 1990, \textbf{36},
  85--108\relax
\mciteBstWouldAddEndPuncttrue
\mciteSetBstMidEndSepPunct{\mcitedefaultmidpunct}
{\mcitedefaultendpunct}{\mcitedefaultseppunct}\relax
\EndOfBibitem
\bibitem[Pignon \emph{et~al.}(1996)Pignon, Magnin, and Piau]{Pignon:1996}
F.~Pignon, A.~Magnin and J.-M. Piau, \emph{J. Rheol.}, 1996, \textbf{40},
  573--587\relax
\mciteBstWouldAddEndPuncttrue
\mciteSetBstMidEndSepPunct{\mcitedefaultmidpunct}
{\mcitedefaultendpunct}{\mcitedefaultseppunct}\relax
\EndOfBibitem
\bibitem[M{\o}ller \emph{et~al.}(2009)M{\o}ller, Fall, Chikkadi, Derks, and
  Bonn]{Moller:2009b}
P.~C.~F. M{\o}ller, A.~Fall, V.~Chikkadi, D.~Derks and D.~Bonn, \emph{Phil.
  Trans. R. Soc. Lond. A}, 2009, \textbf{367}, 5139--5155\relax
\mciteBstWouldAddEndPuncttrue
\mciteSetBstMidEndSepPunct{\mcitedefaultmidpunct}
{\mcitedefaultendpunct}{\mcitedefaultseppunct}\relax
\EndOfBibitem
\bibitem[Coussot and Ovarlez(2010)]{Coussot:2010}
P.~Coussot and G.~Ovarlez, \emph{Eur. Phys. J. E}, 2010, \textbf{33},
  183--188\relax
\mciteBstWouldAddEndPuncttrue
\mciteSetBstMidEndSepPunct{\mcitedefaultmidpunct}
{\mcitedefaultendpunct}{\mcitedefaultseppunct}\relax
\EndOfBibitem
\bibitem[Weeks(2007)]{Weeks:2007}
E.~R. Weeks, Statistical Physics of Complex Fluids, 2007, pp. 243--255\relax
\mciteBstWouldAddEndPuncttrue
\mciteSetBstMidEndSepPunct{\mcitedefaultmidpunct}
{\mcitedefaultendpunct}{\mcitedefaultseppunct}\relax
\EndOfBibitem
\bibitem[Durian(1995)]{Durian:1995}
D.~J. Durian, \emph{Phys. Rev. Lett.}, 1995, \textbf{75}, 4780--4783\relax
\mciteBstWouldAddEndPuncttrue
\mciteSetBstMidEndSepPunct{\mcitedefaultmidpunct}
{\mcitedefaultendpunct}{\mcitedefaultseppunct}\relax
\EndOfBibitem
\bibitem[Okuzono and Kawasaki(1995)]{Okuzono:1995}
T.~Okuzono and K.~Kawasaki, \emph{Phys. Rev. E}, 1995, \textbf{51},
  1246--1253\relax
\mciteBstWouldAddEndPuncttrue
\mciteSetBstMidEndSepPunct{\mcitedefaultmidpunct}
{\mcitedefaultendpunct}{\mcitedefaultseppunct}\relax
\EndOfBibitem
\bibitem[B{\'e}nito \emph{et~al.}(2008)B{\'e}nito, Bruneau, Colin, Gay, and
  Molino]{Benito:2008}
S.~B{\'e}nito, C.-H. Bruneau, T.~Colin, C.~Gay and F.~Molino, \emph{Eur. Phys.
  J. E}, 2008, \textbf{25}, 225–--251\relax
\mciteBstWouldAddEndPuncttrue
\mciteSetBstMidEndSepPunct{\mcitedefaultmidpunct}
{\mcitedefaultendpunct}{\mcitedefaultseppunct}\relax
\EndOfBibitem
\bibitem[Barry \emph{et~al.}(2010)Barry, Weaire, and Hutzler]{Barry:2010}
J.~D. Barry, D.~Weaire and S.~Hutzler, \emph{Rheol. Acta}, 2010, \textbf{49},
  687–--698\relax
\mciteBstWouldAddEndPuncttrue
\mciteSetBstMidEndSepPunct{\mcitedefaultmidpunct}
{\mcitedefaultendpunct}{\mcitedefaultseppunct}\relax
\EndOfBibitem
\bibitem[Kabla \emph{et~al.}(2007)Kabla, Scheibert, and Debregeas]{Kabla:2007}
A.~Kabla, J.~Scheibert and G.~Debregeas, \emph{J. Fluid Mech.}, 2007,
  \textbf{587}, 45--72\relax
\mciteBstWouldAddEndPuncttrue
\mciteSetBstMidEndSepPunct{\mcitedefaultmidpunct}
{\mcitedefaultendpunct}{\mcitedefaultseppunct}\relax
\EndOfBibitem
\bibitem[Roberts and Barnes(2001)]{Roberts:2001}
G.~P. Roberts and H.~A. Barnes, \emph{Rheol. Acta}, 2001, \textbf{40},
  499--503\relax
\mciteBstWouldAddEndPuncttrue
\mciteSetBstMidEndSepPunct{\mcitedefaultmidpunct}
{\mcitedefaultendpunct}{\mcitedefaultseppunct}\relax
\EndOfBibitem
\bibitem[Ketz \emph{et~al.}(1988)Ketz, Prud'homme, and Graessley]{Ketz:1988}
R.~J. Ketz, R.~K. Prud'homme and W.~W. Graessley, \emph{Rheol. Acta}, 1988,
  \textbf{27}, 531--539\relax
\mciteBstWouldAddEndPuncttrue
\mciteSetBstMidEndSepPunct{\mcitedefaultmidpunct}
{\mcitedefaultendpunct}{\mcitedefaultseppunct}\relax
\EndOfBibitem
\bibitem[Kim \emph{et~al.}(2003)Kim, Song, Lee, and Park]{Kim:2003}
J.-Y. Kim, J.-Y. Song, E.-J. Lee and S.-K. Park, \emph{Colloid Polym. Sci.},
  2003, \textbf{281}, 614--623\relax
\mciteBstWouldAddEndPuncttrue
\mciteSetBstMidEndSepPunct{\mcitedefaultmidpunct}
{\mcitedefaultendpunct}{\mcitedefaultseppunct}\relax
\EndOfBibitem
\bibitem[Oppong \emph{et~al.}(2006)Oppong, Rubatat, Bailey, Frisken, and
  de~Bruyn]{Oppong:2006}
F.~K. Oppong, L.~Rubatat, A.~E. Bailey, B.~J. Frisken and J.~R. de~Bruyn,
  \emph{Phys. Rev. E}, 2006, \textbf{73}, 041405\relax
\mciteBstWouldAddEndPuncttrue
\mciteSetBstMidEndSepPunct{\mcitedefaultmidpunct}
{\mcitedefaultendpunct}{\mcitedefaultseppunct}\relax
\EndOfBibitem
\bibitem[Lee \emph{et~al.}(2011)Lee, Gutowski, Bailey, Rubatat, de~Bruyn, and
  Frisken]{Lee:2011}
D.~Lee, I.~A. Gutowski, A.~E. Bailey, L.~Rubatat, J.~R. de~Bruyn and B.~J.
  Frisken, \emph{Phys. Rev. E}, 2011, \textbf{83}, 031401\relax
\mciteBstWouldAddEndPuncttrue
\mciteSetBstMidEndSepPunct{\mcitedefaultmidpunct}
{\mcitedefaultendpunct}{\mcitedefaultseppunct}\relax
\EndOfBibitem
\bibitem[Piau(2007)]{Piau:2007}
J.~M. Piau, \emph{J. Non-Newtonian Fluid Mech.}, 2007, \textbf{144},
  1--29\relax
\mciteBstWouldAddEndPuncttrue
\mciteSetBstMidEndSepPunct{\mcitedefaultmidpunct}
{\mcitedefaultendpunct}{\mcitedefaultseppunct}\relax
\EndOfBibitem
\bibitem[Divoux \emph{et~al.}(2010)Divoux, Tamarii, Barentin, and
  Manneville]{Divoux:2010}
T.~Divoux, D.~Tamarii, C.~Barentin and S.~Manneville, \emph{Phys. Rev. Lett.},
  2010, \textbf{104}, 208301\relax
\mciteBstWouldAddEndPuncttrue
\mciteSetBstMidEndSepPunct{\mcitedefaultmidpunct}
{\mcitedefaultendpunct}{\mcitedefaultseppunct}\relax
\EndOfBibitem
\bibitem[Curran \emph{et~al.}(2002)Curran, Hayes, Afacan, Williams, and
  Tanguy]{Curran:2002}
S.~Curran, R.~E. Hayes, A.~Afacan, M.~Williams and P.~Tanguy, \emph{J. Food
  Sci.}, 2002, \textbf{67}, 176--180\relax
\mciteBstWouldAddEndPuncttrue
\mciteSetBstMidEndSepPunct{\mcitedefaultmidpunct}
{\mcitedefaultendpunct}{\mcitedefaultseppunct}\relax
\EndOfBibitem
\bibitem[Baudonnet \emph{et~al.}(2004)Baudonnet, Grossiord, and
  Rodriguez]{Baudonnet:2004}
L.~Baudonnet, J.-L. Grossiord and F.~Rodriguez, \emph{J. Dispersion Sci.
  Technol.}, 2004, \textbf{25}, 183--192\relax
\mciteBstWouldAddEndPuncttrue
\mciteSetBstMidEndSepPunct{\mcitedefaultmidpunct}
{\mcitedefaultendpunct}{\mcitedefaultseppunct}\relax
\EndOfBibitem
\bibitem[Meeker \emph{et~al.}(2004)Meeker, Bonnecaze, and
  Cloitre]{Meeker:2004a}
S.~P. Meeker, R.~T. Bonnecaze and M.~Cloitre, \emph{Phys. Rev. Lett.}, 2004,
  \textbf{92}, 198302\relax
\mciteBstWouldAddEndPuncttrue
\mciteSetBstMidEndSepPunct{\mcitedefaultmidpunct}
{\mcitedefaultendpunct}{\mcitedefaultseppunct}\relax
\EndOfBibitem
\bibitem[Meeker \emph{et~al.}(2004)Meeker, Bonnecaze, and
  Cloitre]{Meeker:2004b}
S.~P. Meeker, R.~T. Bonnecaze and M.~Cloitre, \emph{J. Rheol.}, 2004,
  \textbf{48}, 1295--1320\relax
\mciteBstWouldAddEndPuncttrue
\mciteSetBstMidEndSepPunct{\mcitedefaultmidpunct}
{\mcitedefaultendpunct}{\mcitedefaultseppunct}\relax
\EndOfBibitem
\bibitem[Manneville \emph{et~al.}(2004)Manneville, B{\'e}cu, and
  Colin]{Manneville:2004a}
S.~Manneville, L.~B{\'e}cu and A.~Colin, \emph{Eur. Phys. J. AP}, 2004,
  \textbf{28}, 361--373\relax
\mciteBstWouldAddEndPuncttrue
\mciteSetBstMidEndSepPunct{\mcitedefaultmidpunct}
{\mcitedefaultendpunct}{\mcitedefaultseppunct}\relax
\EndOfBibitem
\bibitem[Borrega(2000)]{Borrega:2000}
R.~Borrega, \emph{PhD thesis}, Universit{\'e} Paris VI, 2000\relax
\mciteBstWouldAddEndPuncttrue
\mciteSetBstMidEndSepPunct{\mcitedefaultmidpunct}
{\mcitedefaultendpunct}{\mcitedefaultseppunct}\relax
\EndOfBibitem
\bibitem[Benmouffok-Benbelkacem \emph{et~al.}(2010)Benmouffok-Benbelkacem,
  Caton, Baravian, and Skali-Lami]{Benmouffok:2010}
G.~Benmouffok-Benbelkacem, F.~Caton, C.~Baravian and S.~Skali-Lami,
  \emph{Rheol. Acta}, 2010, \textbf{49}, 305--314\relax
\mciteBstWouldAddEndPuncttrue
\mciteSetBstMidEndSepPunct{\mcitedefaultmidpunct}
{\mcitedefaultendpunct}{\mcitedefaultseppunct}\relax
\EndOfBibitem
\bibitem[Cohen-Addad \emph{et~al.}(1998)Cohen-Addad, Hoballah, and
  H\"ohler]{Cohen-Addad:1998}
S.~Cohen-Addad, H.~Hoballah and R.~H\"ohler, \emph{Phys. Rev. E}, 1998,
  \textbf{57}, 6897--6901\relax
\mciteBstWouldAddEndPuncttrue
\mciteSetBstMidEndSepPunct{\mcitedefaultmidpunct}
{\mcitedefaultendpunct}{\mcitedefaultseppunct}\relax
\EndOfBibitem
\bibitem[Gopal and Durian(2003)]{Gopal:2003}
A.~Gopal and D.~Durian, \emph{Phys. Rev. Lett.}, 2003, \textbf{91},
  188303\relax
\mciteBstWouldAddEndPuncttrue
\mciteSetBstMidEndSepPunct{\mcitedefaultmidpunct}
{\mcitedefaultendpunct}{\mcitedefaultseppunct}\relax
\EndOfBibitem
\bibitem[Mason and Weitz(1995)]{Mason:1995a}
T.~G. Mason and D.~A. Weitz, \emph{Phys. Rev. Lett.}, 1995, \textbf{74},
  1250--1253\relax
\mciteBstWouldAddEndPuncttrue
\mciteSetBstMidEndSepPunct{\mcitedefaultmidpunct}
{\mcitedefaultendpunct}{\mcitedefaultseppunct}\relax
\EndOfBibitem
\bibitem[Mason \emph{et~al.}(1995)Mason, Bibette, and Weitz]{Mason:1995c}
T.~G. Mason, J.~Bibette and D.~A. Weitz, \emph{Phys. Rev. Lett.}, 1995,
  \textbf{75}, 2051--2054\relax
\mciteBstWouldAddEndPuncttrue
\mciteSetBstMidEndSepPunct{\mcitedefaultmidpunct}
{\mcitedefaultendpunct}{\mcitedefaultseppunct}\relax
\EndOfBibitem
\bibitem[H{\'e}braud \emph{et~al.}(2000)H{\'e}braud, Lequeux, and
  Palierne]{Hebraud:2000}
P.~H{\'e}braud, F.~Lequeux and J.-F. Palierne, \emph{Langmuir}, 2000,
  \textbf{16}, 8296--8299\relax
\mciteBstWouldAddEndPuncttrue
\mciteSetBstMidEndSepPunct{\mcitedefaultmidpunct}
{\mcitedefaultendpunct}{\mcitedefaultseppunct}\relax
\EndOfBibitem
\bibitem[Liu \emph{et~al.}(1996)Liu, Ramaswamy, Mason, Gang, and
  Weitz]{Liu:1996b}
A.~J. Liu, S.~Ramaswamy, T.~Mason, H.~Gang and D.~Weitz, \emph{Phys. Rev.
  Lett.}, 1996, \textbf{76}, 3017--3020\relax
\mciteBstWouldAddEndPuncttrue
\mciteSetBstMidEndSepPunct{\mcitedefaultmidpunct}
{\mcitedefaultendpunct}{\mcitedefaultseppunct}\relax
\EndOfBibitem
\bibitem[Sollich \emph{et~al.}(1997)Sollich, Lequeux, H\'ebraud, and
  Cates]{Sollich:1997}
P.~Sollich, F.~Lequeux, P.~H\'ebraud and M.~E. Cates, \emph{Phys. Rev. Lett.},
  1997, \textbf{78}, 2020--2023\relax
\mciteBstWouldAddEndPuncttrue
\mciteSetBstMidEndSepPunct{\mcitedefaultmidpunct}
{\mcitedefaultendpunct}{\mcitedefaultseppunct}\relax
\EndOfBibitem
\bibitem[Divoux \emph{et~al.}()Divoux, Barentin, and Manneville]{Divoux:2011pp}
T.~Divoux, C.~Barentin and S.~Manneville, E-print cond-mat/1012.0693\relax
\mciteBstWouldAddEndPuncttrue
\mciteSetBstMidEndSepPunct{\mcitedefaultmidpunct}
{\mcitedefaultendpunct}{\mcitedefaultseppunct}\relax
\EndOfBibitem
\bibitem[Barnes(1995)]{Barnes:1995}
H.~A. Barnes, \emph{J. Non-Newtonian Fluid Mech.}, 1995, \textbf{56},
  221--251\relax
\mciteBstWouldAddEndPuncttrue
\mciteSetBstMidEndSepPunct{\mcitedefaultmidpunct}
{\mcitedefaultendpunct}{\mcitedefaultseppunct}\relax
\EndOfBibitem
\bibitem[Buscall(2010)]{Buscall:2010}
R.~Buscall, \emph{J. Rheol.}, 2010, \textbf{54}, 1177--1183\relax
\mciteBstWouldAddEndPuncttrue
\mciteSetBstMidEndSepPunct{\mcitedefaultmidpunct}
{\mcitedefaultendpunct}{\mcitedefaultseppunct}\relax
\EndOfBibitem
\bibitem[Gibaud \emph{et~al.}(2008)Gibaud, Barentin, and
  Manneville]{Gibaud:2008}
T.~Gibaud, C.~Barentin and S.~Manneville, \emph{Phys. Rev. Lett.}, 2008,
  \textbf{101}, 258302\relax
\mciteBstWouldAddEndPuncttrue
\mciteSetBstMidEndSepPunct{\mcitedefaultmidpunct}
{\mcitedefaultendpunct}{\mcitedefaultseppunct}\relax
\EndOfBibitem
\bibitem[Lettinga and Manneville(2009)]{Lettinga:2009}
M.~P. Lettinga and S.~Manneville, \emph{Phys. Rev. Lett.}, 2009, \textbf{103},
  248302\relax
\mciteBstWouldAddEndPuncttrue
\mciteSetBstMidEndSepPunct{\mcitedefaultmidpunct}
{\mcitedefaultendpunct}{\mcitedefaultseppunct}\relax
\EndOfBibitem
\bibitem[Tanguy \emph{et~al.}(2006)Tanguy, Leonforte, and Barrat]{Tanguy:2006}
A.~Tanguy, F.~Leonforte and J.-L. Barrat, \emph{Eur. Phys. J. E}, 2006,
  \textbf{20}, 355--364\relax
\mciteBstWouldAddEndPuncttrue
\mciteSetBstMidEndSepPunct{\mcitedefaultmidpunct}
{\mcitedefaultendpunct}{\mcitedefaultseppunct}\relax
\EndOfBibitem
\bibitem[Tsamados(2010)]{Tsamados:2010}
M.~Tsamados, \emph{Eur. Phys. J. E}, 2010, \textbf{32}, 165--181\relax
\mciteBstWouldAddEndPuncttrue
\mciteSetBstMidEndSepPunct{\mcitedefaultmidpunct}
{\mcitedefaultendpunct}{\mcitedefaultseppunct}\relax
\EndOfBibitem
\end{mcitethebibliography}
\bibliographystyle{rsc} 

\end{document}